\documentstyle[aasms4]{article}

\begin{document}

\title{The PDS starburst galaxies}

\author{Roger Coziol\altaffilmark{1,}\altaffilmark{2}, Carlos A. O. Torres\altaffilmark{1}, Germano
R. Quast\altaffilmark{1}, Thierry Contini\altaffilmark{3}}
\and
\author{Emmanuel Davoust\altaffilmark{4}}

\altaffiltext{1}{Laborat\'orio Nacional de Astrof\'{\i}sica - LNA/CNPq,
CP 21, 37500--000 Itajub\'a, MG, Brazil} 
\altaffiltext{2}{Present address: Observat\'orio Nacional,  
CEP 20921-400, Rio de Janeiro, Brasil}
\altaffiltext{3}{School of Physics \& Astronomy, Tel Aviv University, 69978 Tel Aviv, Israel}
\altaffiltext{4}{Observatoire Midi-Pyr\'en\'ees, UMR 5572, F-31400 Toulouse, France}

\begin{abstract}

We discuss the nature of the galaxies found in the 
Pico dos Dias Survey (PDS) for young stellar objects. 
The PDS galaxies were selected from the IRAS Point Source 
catalog. They have flux density of moderate or high quality at 12, 25 
and 60 $\mu$m and spectral indices in the ranges $-3.00 \leq \alpha(25,12)
\leq +0.35$ and  $-2.50 \leq \alpha(60,25) \leq +0.85$.
These criteria allowed the detection of 382 galaxies, which are 
a mixture of starburst and Seyfert galaxies. 
Most of the PDS Seyferts are included in the catalog of warm IRAS sources 
by de Grijp et al.\ (1987).  
The remaining galaxies constitute a homogeneous sample of luminous 
($\log({\rm L}_{\rm B}/{\rm L}_\odot) = 9.9 \pm 0.4$) starburst galaxies,
67\% of which were not recognized as such before.

The starburst nature of the PDS galaxies is established by comparing their 
L$_{\rm IR}$/L$_{\rm B}$ ratios and IRAS colors with a sample of emission
line galaxies from the literature already classified as starburst galaxies.
The starburst galaxies show an excess of FIR luminosity and their IRAS colors are 
significantly different from those of Seyfert galaxies -- 99\%
of the starburst galaxies in our sample have a spectral index $\alpha(60,25) < -1.9$. 
As opposed to Seyfert galaxies, very few PDS starbursts are detected in X-rays.

In the infrared, the starburst galaxies form a continuous sequence with normal galaxies.
But they generally can be distinguished from normal galaxies by their spectral 
index $\alpha(60,25) > -2.5$.
This color cut--off also marks a change in the dominant morphologies of the galaxies: 
the normal IRAS galaxies are preferentially late--type spirals (Sb and later), 
while the starbursts are more numerous among early--type spirals (earlier than Sbc). 
This preference of starbursts for early--type spirals is not new, but a trait 
of the massive Starburst Nucleus Galaxies (Coziol et al.\ 1997a).
Like in other SBNG samples, the PDS starbursts show no preference for barred
galaxies.

No difference is found between the starbursts detected in the FIR 
and those detected on the basis of UV excess. 
The PDS starburst galaxies represent the FIR luminous branch of 
the UV-bright starburst nucleus galaxies, with mean FIR luminosity $\log({\rm L}_{\rm IR}/{\rm L}_\odot) 
= 10.3 \pm 0.5$ and redshifts smaller than 0.1. They form a complete
sample limited in flux in the FIR at $2\times10^{-10}$ erg cm$^{-2}$ s$^{-1}$. 
\end{abstract}

\keywords{surveys -- galaxies: starburst -- galaxies: Seyfert -- infrared: galaxies -- X-rays: galaxies}

\section{Introduction} 

Since their discovery a few decades ago, starburst galaxies 
have been a constantly growing field of research. This increasing
interest is due to the fact that what was once considered as a peculiar phenomenon affecting 
only a small fraction of the local population of galaxies turns out to be of 
a more general nature, involving mechanisms by which galaxies form 
and evolve (Larson 1990; Kennicutt 1990; van den Bergh et al.\ 1996;
Coziol et al.\ 1997b, 1998; Driver et al.\ 1998).
Yet, our present knowledge on the properties of starburst galaxies 
rests on the analyses of a rather limited number of data sets and 
fundamental questions on their nature still remain to be answered. For example, is there a relation  
between the small-mass and metal poor \ion{H}{2} galaxies and the
more massive and chemically evolved Starburst Nucleus Galaxies (SBNGs)? 
What is the difference between starburst galaxies which are
UV-bright and those which are selected based on a strong emission 
in the far infrared (FIR)? Is there a relation between Active Galactic Nuclei (AGN) 
and starburst activity? Are all starbursts efficient X-rays emitters? 
What is the dominant morphology of starburst galaxies?

In this context, we have judged useful to present this new homogeneous 
sample of relatively nearby and luminous starburst galaxies selected 
in the FIR: the PDS (which stands for Pico dos Dias Survey) starburst galaxies. 
The fact that starbursts are easily detected in the FIR is 
not surprising (Meadows et al 1990; Allen et al.\ 1991; Contini et al.\ 1998). 
In their selected sample of IRAS galaxies, for example, Allen et al.\ (1991) found 90\% 
starburst and 10\% Seyfert galaxies. Similar ratios were found in the
Contini et al.\ (1998) sample of Markarian UV-bright galaxies.
In the Montreal Blue Galaxy (MBG) survey,  85\% of the UV-bright starburst 
galaxies are also IRAS sources (Coziol et al.\ 1997a). 
Interestingly, the percentage of AGNs seems to increase with the FIR luminosity.  
In their sample of Luminous Infrared Galaxies (LIGs), Veilleux et al.\ (1995)
found 41\% Seyfert galaxies and 59\% starbursts.  
For the PDS galaxies, the discovery ratios are 38\% AGNs for 62\% starbursts. 

The organization of this paper is as follows. In section~\ref{PDSGAL}, we present the 
list of the PDS starburst galaxies. In section~\ref{STBNAT}, the starburst nature of these 
galaxies is established by comparing their IRAS and optical characteristics 
with those of normal and starburst galaxies taken from the literature. 
The IRAS color--color diagrams of the PDS starbursts are compared 
to those of the AGNs and two criteria are proposed to distinguish between 
these two types of activity in galaxies. The application of these criteria 
allows us to identify 9 misclassified Seyfert galaxies. In section~\ref{PDSPROP}, 
some of the characteristics of the PDS starbursts are examined. A summary of our 
results is given in section~\ref{CONCLU}. 

\section{The sample of PDS galaxies}
\label{PDSGAL}

The Pico dos Dias survey (PDS) is a systematic search performed at the Observat\'orio 
do Pico dos Dias (OPD -- operated by the Laborat\'orio Nacional de Astrof\'{\i}sica (LNA), 
Conselho Nacional de Desenvolvimento Cient\'{\i}fico e Tecnol\'ogico (CNPq))
to discover young stellar objects from their FIR properties (Gregorio-Hetem et al.\ 1992; 
Torres et al.\ 1995). The candidates, selected from the IRAS Point Source 
Catalog (IPSC), have flux density qualities 2 or 3 at 12, 25 
and 60 $\mu$m, and spectral indices in the ranges $-3.00 \leq \alpha(25,12)
\leq +0.35$ and  $-2.50 \leq \alpha(60,25) \leq +0.85$  (Torres \& Quast 1995),
where the spectral indices are defined as $\alpha(\lambda 1,\lambda 2) = 
\log(S_{\lambda1}/ S_{\lambda2})/\log( \lambda2/\lambda1)$, and 
$S_{\lambda1}$ is the flux in Jansky at wavelength $\lambda1$.
The appearance of all the young stellar candidates was examined on the 
Digitized Sky Survey plates and the galaxy-like objects were placed on a 
separate list, which we call the PDS galaxy list. 
Some star-like AGNs were also included in this list after establishing their
nature from the V\'eron--Cetty \& V\'eron (1996) compilation.
The goal of the present paper is to determine the nature of all 
the objects in the PDS galaxy list. 

Initially, the PDS galaxies were composed of 388 objects. 
An entry for all these sources was found in 
NED\footnote{The NASA/IPAC Extragalactic Database.}. Two objects turned out to 
be planetary nebulae and 3 others were identified with \ion{H}{2} regions in M33 and 
M101. The luminous quasar 3C 273 was also detected. 
After removing these 6 objects, we were left with 382 galaxies. 

There are 122 AGNs in our list, distributed among 4 LINERs, 76 Seyfert 2 (Sy2) and 42 
Seyfert 1 (Sy1). The fact that there are so few LINERs is intriguing, 
considering the high volume density of these objects in the nearby 
Universe ($\sim 30$\% of the local luminous galaxies; Heckman 1980; Ho, Filippenko 
\& Sargent 1997). We note that 84\% of the PDS AGNs are included in 
the warm IRAS source catalog of de Grijp et al.\ (1987); this suggests that 
they used similar selection criteria in the 
FIR as ours. Because de Grijp et al.\ were only interested in AGNs, they 
applied an arbitrary cut-off to the color $\alpha$(60,25) to separate them 
from normal galaxies. In the PDS, the limit applied to this color is slightly 
below this cut-off, allowing the inclusion of numerous non-AGN galaxies. 
However, we will show that these galaxies are not normal but mostly 
starbursts. 

The 260 PDS non-AGN galaxies in our list are either already known starburst 
galaxies (33\%) or unclassified galaxies that we consider to be starbursts.  
It is noteworthy that 67\% of the PDS starburst galaxies were not recognized 
as such before.                      

To facilitate our analysis, we retained only the galaxies which 
have fluxes with high or intermediate qualities in the IRAS Faint Source 
Catalog (IFSC). We rejected 14 galaxies of low quality flux and 46 others because 
they have a Galactic latitude $ |b| < 10^{\circ}$, and consequently do not appear in 
the IFSC. 200 galaxies were thus left in our sample. 
 
In Table 1 we present the 200 PDS starburst candidates. Column 1 gives their 
IRAS name. A capital X at the end of the name identifies the galaxy as an X-ray
source, according to the ROSAT catalog (Voges et al.\ 1996).
Column 2 gives another acronym from NED.    
Other properties also taken from NED are: the 
1950 coordinates of the galaxy (columns 3 and 4), the redshift (column 4), 
the B magnitude (column 6) and the morphology (column 7). We note 
that our list contains two of the most famous starbursts: M82 and NGC 7714. 

In Table 2, we give the optical and FIR characteristics of our sample of 
galaxies. Columns 2 and 3 correspond to the absolute magnitude and B 
luminosity, which were determined using the magnitudes quoted in Table 1 and 
assuming H$_0 = 75$ km s$^{-1}$ Mpc$^{-1}$. No correction for Galactic 
reddening was applied. The infrared luminosity in column 4 was determined 
from the relation: $\log({\rm L}_{\rm IR}) = \log({\rm F}_{\rm IR}) + 2 \log[z(z+1)] + 57.28$, 
where $z$ is the redshift and ${\rm F}_{\rm IR} = 1.26 \times 10^{-
11} (2.58 S_{60} + S_{100})$ erg cm$^{-2}$ s$^{-1}$ (Londsdale et al.\ 1985). 
The mean FIR luminosity of the PDS starbursts is $2 \times 10^{10} {\rm L}_{\odot}$, 
slightly less than the mean luminosity of the LIGs 
studied by Veilleux et al.\ (1995).
In columns 5, 6, and 7 we give the IRAS spectral indices. In general,
the differences between the spectral indices determined using the IFSC
or IPSC fluxes are marginal. 
The quality indices in the four IRAS bands are given in column 8.     

An interesting characteristic of our sample is the different fractions of 
galaxies with different activity types detected in X-rays: 2 of the 4 LINERs (50\%) were 
detected, as compared to 71\% of the Sy1, 22\% of the Sy2 and only 
4\% of the starbursts. These differences between the starbursts and the two Seyfert galaxies 
are highly significant considering that in our sample the Sy2 
are almost twice as numerous as the Sy1 and 
the starbursts still more numerous than the Sy2.  

\section{The starburst nature of the PDS galaxies}
\label{STBNAT}

We define a starburst galaxy as an emission--line galaxy where
the source of ionization of the gas is of stellar origin, as determined from 
standard emission line ratio diagnostic diagrams (Baldwin, Phillips \& Terlevich 1981; 
Veilleux \& Osterbrock 1987). This first criterion eliminates the Seyfert galaxies, 
but also the LINERs whose source of ionization is not clearly defined.
The starburst galaxies have H$\alpha$ luminosities from 10$^{39}$ up to a few 
10$^{42}$ erg s$^{-1}$ (Balzano 1983; Coziol 1996), implying unusually high present star 
formation rates. Obviously, these galaxies cannot sustain
such elevated star formation rates for a very long time (hence the name starburst; 
Weedman et al.\ 1981), unless they are replenished in gas, following some kind of interaction with
another galaxy or with its environment (Huchra 1977; Taylor et al.\ 1996), or if the star formation is 
regulated by internal processes, probably related to supernovae 
feedback (Searle \& Sargent 1972; Gerola, Seiden \& Schulman, 1980; Kr\"{u}gel \& Tutukov 1993).  
From the University of Michigan (UM) survey, Salzer et al.\ (1989) have shown that 
the starbursts present a variety of types related to their morphologies. 
Based on similar characteristics, we can regroup these types in two broad categories (Coziol et al.\ 1994): the \ion{H}{2} galaxies which are small mass and metal poor galaxies and the Starburst Nucleus Galaxies (SBNGs) which are more massive and chemically evolved. 

Starburst galaxies are usually detected by the presence of emission--lines in their spectra (using
the objective--prism technique) or by a UV color excess (using   
the multiple filters technique). The two methods yield similar results (Coziol et al.\ 1993), the
only difference being that objective--prism surveys are biased against SBNGs while 
multiple colors surveys are biased against \ion{H}{2} galaxies (Coziol 1996). Starburst galaxies  
detected in the FIR are generally similar to the SBNGs detected in the optical
(Allen et al.\ 1991; Veilleux et al.\ 1995). However, at the extreme end of the FIR
range one finds a new type of starburst galaxy: 
the ultra Luminous Infrared Galaxies (uLIGs), which emit most of their 
energy in the FIR (Sanders et al.\ 1988). One important property distinguishes 
the uLIGs from the other two types of starbursts: while all the uLIGs are cases 
of massive galaxies in interaction (Mirabel \& Duc 1993), only
a small fraction ($\sim 1/4$) of the SBNGs and \ion{H}{2} galaxies clearly are
(Telles \& Terlevich 1995; Coziol et al.\ 1997).                      

In what follows, we determine the starburst nature of the galaxies in our 
sample from their FIR luminosity and color. Our method is semi--empirical in nature. We
use simple models to interpret the L$_{\rm IR}$/L$_{\rm 
B}$ ratios (Coziol et al.\ 1996) and IRAS colors (Sekiguchi 1987) 
of the galaxies
and compare the characteristics of the PDS galaxies with those of 
different samples of normal and starburst galaxies taken from the literature. 
The sample of normal galaxies is composed of galaxies from the 
list of Stauffer (1982), Kennicutt (1983) and 
Hogg et al.\ (1993), from which we eliminated, using their identification 
in NED, all the known Seyferts, interacting, UV-bright and 
starburst galaxies (such as UM, Arp, Markarian, Kiso, MBG or 
Zwicky-like galaxies). The sample of starburst galaxies is composed of three
samples representing the different types of starbursts:  
(a) SBNGs from the UM survey (Salzer et al.\ 1989), the Balzano's sample of
Markarian galaxies (1983) and the MBG survey (Coziol et al.\ 1993, 1994, 1997a, 1997b);  
(b) \ion{H}{2} galaxies  from the  catalog of Terlevich et al.\ (1991) and from the Calan--Tololo survey (Pe\~na et al.\ 1991); 
(c) Luminous Infrared Galaxies (LIGs), as observed by Veilleux et al.\ (1995), which most luminous members are
similar to the uLIGs. 

\subsection{The FIR luminosity excess in starbursts}

In Figure 1, we present the diagram of L$_{\rm IR}$ as a function of L$_{\rm 
B}$. Coziol (1996) showed that starburst galaxies present a typical 
excess of FIR luminosity as compared to normal spiral galaxies. In Figure 1a, 
this phenomenon is illustrated by comparing the luminosities of the PDS 
starbursts with the mean values found for different samples of galaxies, as determined
in Coziol (1996). For comparison purposes, we distinguished between 
early-type (earlier than Sbc) and late-type galaxies (Sbc and later).
It can be seen that most of the PDS galaxies show a ratio L$_{\rm IR}$/L$_{\rm B}$ higher than 1, 
similar to the SBNGs. 

In Figure 1b, which concerns the PDS Seyfert galaxies, 
we also show the mean values found by Roberts \& Haynes (1994) 
for normal galaxies with different morphologies. This sample is probably more
representative of normal galaxies than the samples of Stauffer, Hogg et al.\ or Kennicutt, because it
is not biased towards emission-line galaxies. Comparing the PDS galaxies
(starbursts and AGNs) with the normal galaxies, it can be seen that  
the former have normal blue luminosities.   
Like the SBNGs, their mean absolute B magnitude M$_{\rm B} = -20.0 \pm 1.0$ suggests that 
they generally are massive galaxies. 
This result is consistent with the fact that small-mass 
starburst galaxies (like the \ion{H}{2} 
galaxies for example) are generally deficient in dust (Coziol 1996).  

Although the FIR-luminosity-excess criterion enables one 
to distinguish starbursts from normal galaxies, it does not allow to separate
them from AGNs.  Indeed, in Figure 1b we see that, in general, the Seyfert galaxies
show L$_{\rm IR}$/L$_{\rm B}$ ratios comparable to those of SBNGs. This is a characteristic of Seyfert 
galaxies (Coziol 1996). The nature of the FIR excess in these galaxies 
is ambiguous, because the contribution of the active nucleus to the FIR 
luminosity is not well determined. If this contribution were negligible, then 
the excess of FIR luminosity would imply high star forming rates. 

\subsection{A model for the typical FIR colors of galaxies} 

In Figure 2a, we show the diagram of the spectral indices $\alpha$(60,25) versus 
$\alpha$(100,60) for the sample of normal galaxies, as defined above. It can be seen 
that normal galaxies occupy only a small region of this diagram.
There also seems to be no difference in colors between the
normal early--type galaxies and the late--types ones (see
Sauvage \& Thuan 1994 for a discussion of this phenomenon in normal galaxies). 

In Figure 2b, the same diagram is shown for the samples of starburst 
galaxies. The FIR colors of the starbursts are clearly offset from 
those of normal galaxies, but a significant region of overlap exists.  
This result reminds one of the observation made by Huchra (1977) about the 
Markarian galaxies, namely that starbursts are not a new class of 
objects but rather a subset of normal galaxies. This seems to be true in the FIR as well.

To interpret the data, we use a two-blackbody model composed of a 
cirrus-like, cold component with a temperature $\sim 27$K, added to a 
hot component, associated with a burst of star formation (Sekiguchi 1987). In 
Figure 2, the temperature of the hot component varies from 60K to 100K. The 
numbers on the grid indicate the fractional contribution of the hot component to 
the total FIR luminosity. For the sample of normal galaxies, the contribution 
of the hot component varies between 0.6 and 0.9 and the temperature varies 
between 65K and 80K. In the starbursts, the contribution from the hot component is 
generally higher than 0.8 and the temperature is almost always higher than 
70K. The interpretation of this difference is straightforward (Sekiguchi 1987; Coziol 1996): 
the higher star formation rates in starburst galaxies simply
increases the quantity of hot dust. 

It is instructive to study the location of the various types of 
starbursts in Figure 2b.  The small dispersion for the colors of the SBNGs, 
for example, suggests similar characteristics for the bursts (similar 
intensities, ages or dust contents). The larger dispersion for the LIGs, on 
the other hand, suggests a more heterogeneous group.  This 
difference between the two types of starbursts may be explained 
in part by different levels of extinction, which is generally higher in the LIGs than in 
the optically selected SBNGs (Veilleux et al.\ 1995). In Figure 2b for instance, some LIGs 
have colors similar to a pure blackbody. In these galaxies most of the 
light may be absorbed by dust and reemitted in the FIR. 
The fact, on the other hand, that many UV-bright SBNGs are also FIR emitters suggests 
that the dust distribution in these galaxies is rather patchy (Calzetti et al.\ 1996). 

On average, the LIGs show a higher hot-component contribution and/or a higher 
dust temperature than the SBNGs. The LIGs, therefore, may have higher star formation rates 
or younger bursts than the SBNGs (Coziol 1996). But, some may
also hide an AGN (Sanders et al.\ 1988). 

For the \ion{H}{2} galaxies,
the high dust temperatures suggested by the model are probably better explained by 
their low metallicities and young ages. A young starburst contains a
larger number of massive stars than a more evolved one, and a metal-poor ionized 
gas reaches higher temperatures than a metal-rich one.   

\subsection{Discrimination between normal, starburst and Seyfert galaxies using FIR colors}

In Figure 3a, we now apply our model to the PDS starburst galaxies.
We deduce that the PDS starburst galaxies have IRAS colors which are typical of SBNGs. 
This conclusion is consistent with our previous classification 
based on the ratio L$_{\rm IR}$/L$_{\rm B}$. 

In Figure 3a, we distinguish between the PDS starbursts detected 
in the UV, the galaxies assumed to be in interaction (the Arp galaxies) and the 
pure IRAS galaxies (that is, the FIR galaxies with no other special 
known characteristics). In general, the 
colors of the three types of galaxies are similar. Therefore, whatever the 
origin of the bursts in these galaxies, the result in terms of colors 
seems to be the same. 

In order to test how arbitrary our criteria for selecting starbursts 
in the FIR are, we have examined the nature of the galaxies which have a 
IPSC color $\alpha(60,25) < -2.5$. This specific criterion was tested, 
because we noticed that it eliminates a good number of IRAS galaxies in the IPSC.
Using NED, we identified a new sample of 210 IRAS galaxies 
with a IPSC color $\alpha(60,25) < -2.5$. 
Among these galaxies only 23 (11\%) were recognized as
Seyfert galaxies. We call the remaining 187 galaxies the 
IRAS normal galaxies. Indeed, in Figure 2a,  
we can see that the IRAS normal galaxies have colors typical
of normal galaxies. 

In Table 3, the mean B and FIR luminosities and dispersions for the IRAS normal galaxies are compared 
to those of the PDS starburst and Seyfert galaxies. All these galaxies have comparable luminosities.
In column 4 of Table 3, we also give the L$_{\rm IR}$/L$_{\rm B}$ ratios observed
in all these galaxies. 
The IRAS normal galaxies do not show the same excess of FIR luminosity as the PDS starbursts.
A Kolmogorov--Smirnov test allows one to reject, at a level of confidence higher than 99.9\%,
the hypothesis that the two distributions were taken from the same population.  
Therefore, without completely separating them from normal galaxies our color criteria 
allow us to effectively select starburst galaxies. 
  
In Figure 3b, we present the color-color diagram $\alpha$(60,25) vs.  
$\alpha$(100,60) for the PDS Seyfert galaxies. As already noted by de Grijp et 
al. (1987), the FIR spectral indices of Seyfert galaxies are generally flatter than 
those of starbursts. Only a small fraction of the PDS Seyfert 
galaxies in Figure 3b have FIR colors similar to those of the starbursts. 
There seems to be more Sy2 than Sy1 in this situation: about 38\% of 
the Sy2 compared to only 10\% of the Sy1. Above the cut-off defined by de 
Grijp et al.\ ($\alpha$(60,25) $> -1.5$) the Sy2 show a slightly flatter 
$\alpha$(100,60) spectral index than the Sy1. In Table 3, we also see that the Sy2 
show a similar mean excess of FIR luminosity to that of 
the starbursts, while this ratio for
the Sy1 is normal. Comparing them with the IRAS normal galaxies, a 
Kolmogorov--Smirnov test allows to
reject, but only at the 92\% confidence level, the hypothesis that the Sy2 distributions 
comes from the same population as the normal ones.
The same hypothesis simply cannot be rejected for the Sy1.  

The differences in FIR characteristics among 
the various types of galaxies are better observed 
in the diagrams $\alpha$(60,25) vs. $\alpha$(25,12), presented in Figure 
4. In Figure 4a, most of the PDS starbursts have a color $\alpha$(60,25) 
between $-1.9$ and $-2.5$. In Figure 4b, the PDS Sy2 with
a color $\alpha$(60,25) higher than $-1.9$ have a color
$\alpha$(25,12) lower than $-1.5$, while it is the contrary for the Sy1. This last cut-off also seems to 
separate most of the X-ray AGNs from the non X-ray ones. But it is the 
contrary for the starbursts. 

In Figure 4b, we have also placed the IRAS normal galaxies. As can be seen,
the FIR color distribution of the normal galaxies merges with those
of the starburst galaxies. In reality, all these galaxies form a continuous sequence in colors.
We distinguished the starbursts based on the color criterion\footnote{Note
that it is the spectral index as determined using the IFSC fluxes that we now use.}
$\alpha$(60,25) $> -2.5$.
In section~\ref{PDSPROP}, we will show that this color
cut--off also marks a change in the dominant morphological types of the galaxies. 

\subsection{Spectroscopic observations of misclassified AGNs}   

In Figure 4a, 17 of the PDS starbursts show a spectral index $\alpha$(60,25)$ > 
-1.9$. Considering the large difference between the rest of the starbursts and 
the AGNs in this diagram, we wonderered if these 17 starbursts could not be 
misidentified AGNs. Likewise, the fact that many more PDS AGNs than starbursts 
are detected in X-rays (only 11 PDS starbursts are X-ray sources) 
also suggests 
that some of these starbursts are misidentified AGNs. To verify the nature 
of these galaxies, spectroscopic observations were obtained for 12 of the 17 
starbursts with $\alpha$(60,25)$ > -1.9$ and 3 X-ray starbursts. In August 
1997, 2 galaxies (IC2202 and IRAS14454-4343) were observed with the 1.6m 
telescope at the OPD and 2 others (IIZw083 and IRAS19265-4338) were observed  
at the 1.52m telescope at ESO. The remaining starbursts were observed in 
January and March 1998 with the 1m WISE telescope in Israel. 

At the OPD, a Boller \& Chivens spectrograph was used in conjunction with a 
1024 $\times 1024$, SIT, back-illuminated CCD. Two spectra with resolution 
$\sim 2$\ \AA\ were taken using a 600l/mm grating centered alternatively at 
4600 and 6300\ \AA. The slit had a width of $\sim 2.5$ arsec and was aligned 
in position E--W across the center of the galaxies. At the 1.52m of ESO, the data 
were acquired with a Boller \& Chivens spectrograph and a 2048$\times$2048 
Loral, UV flooded CCD. The grating used had a dispersion of 187 \AA/mm 
providing a spectral coverage of $\sim 3000-9000$\ \AA\ and a resolution of 
about 10\ \AA. The slit width covered $\sim 3$ arsec of the center of the 
galaxies and was positioned close to the parallactic angle. At the 1m telescope 
of the Wise Observatory (Israel) a FOSC spectrograph was used with a 
1024$\times$1024 Tektronics CCD. The spectral coverage  was $\sim 3500-7300$\ 
\AA\ and the resolution was about 8\ \AA. The slit width covered $\sim 5$ arsec 
of the center of the galaxies and was also positioned close to the parallactic 
angle. All the spectra were reduced according to standard procedures under IRAF. 
These include bias subtraction, flat-field and distorsion corrections, cosmic 
ray removal, sky subtraction, wavelength and flux calibrations. 

Table 4 gives the ratios of the most prominent emission lines which we  
used to classify the activity types of the galaxies. Although some of these 
galaxies have very strong emission lines, they do not seem to have broad 
components typical of Sy1 galaxies. 
Figure 5 shows one of the diagnostic diagrams which we used for  
our classification. The results are reported in column 7 of Table 4. Note that the 
same classification is obtained using other line ratios. As we suspected, 9 of the 12 PDS starbursts 
with  $\alpha$(60,25)$ > -1.9$ (identified by crosses in Figure 4a)
are misidentified AGNs (Sy2 or LINER). The three remaining starburst galaxies have color 
values at the borderline between those of starbursts and AGNs (in Figure 4a, they 
are identified by a square and occupy the extreme positions at the top of the sequence traced by the 
starbursts). The color difference between the AGNs and the 
starbursts, in Figure 4a, seems remarkably well established since 99\% of the 
PDS starbursts have a spectral index $\alpha$(60,25)$ > -1.9$.  

Following our spectral classification, the starburst nature of the 3 X-ray
galaxies classified as starbursts on the basis of their colors
is confirmed, although NGC 3310 could also be a LINER. The case of NGC 3690 is 
complicated by the fact that this is really two interacting galaxies. It 
is not clear, therefore, what is the origin of the X-ray emission in this 
system. It could come from hot gas between the two interacting galaxies or be related 
only to one of the galaxies. Again we note that 
the western galaxy of this pair shows line ratios at  
the borderline separating starbursts from LINERs.  As a common 
characteristic, the 4 X-ray starbursts have slightly higher excitation levels than 
normal SBNGs. All these galaxies could be of intermediate nature between 
AGN and starburst (V\'eron et al.\ 1997). 
     
\section{Properties of the PDS starbursts}
\label{PDSPROP}

In Figure 1, we have verified that the PDS galaxies are preferentially 
massive galaxies. This result is consistent with the observation that 
small-mass 
starburst galaxies are usually deficient in dust. In the \ion{H}{2} galaxies  
the dust could have been easily ejected into the intergalactic medium
(with a good part of the metals) by strong starburst winds. Alternatively, these galaxies may
also be too young and have 
not had enough time to produce a sufficient amount of 
dust to be visible in the FIR. Being massive, the PDS galaxies are more  
similar to SBNGs. In this section, we will therefore continue our 
discussion by comparing some of the characteristics of the PDS starburst 
galaxies with those of the SBNGs.  

In Figure 6, we present the morphologies of the PDS starbursts compared to 
those of the Markarian galaxies and the SBNGs of Balzano (1983). We can see 
that the distribution of the morphologies of the PDS galaxies is 
intermediate between those of these two samples. 
In particular, the PDS galaxies seem to contain a relatively high number of 
early-type spirals (Sb and earlier). This is a common trend of SBNG samples
(Coziol et al.\ 1997a). In Figure 6, we verify that this trend in favor of 
early-type spirals is not observed in the sample of IRAS normal galaxies. In 
this case, the fraction of early-type galaxies falls drastically against that 
of the late-type ones. 

We can also see in Figure 6 that the fraction of barred PDS starbursts  
is similar to that observed in other samples of SBNGs. This fraction is 
high, but not particularly so, if we judge from the high fraction of barred 
galaxies also observed in the IRAS normal galaxy sample. Like in other samples of 
SBNGs (Coziol et al.\ 1997a; Continni et al.\ 1998) the fraction of PDS galaxies with a bar seems to 
increase towards late-type galaxies.  The PDS galaxies show no evidence 
to be preferably barred. 

In Figure 7, we compare the FIR luminosities and the redshifts of the PDS 
galaxies with those of the UV-bright MBGs and IRAS galaxies selected by Allen 
et al.\ (1991). Coziol et al.\ (1997a) showed that the MBGs are the 
equivalent of the IRAS luminous galaxies at lower redshifts. The PDS 
starbursts, on the other hand, are observed in similar redshift ranges (from 0 
to 0.1) as the MBGs.  These two samples are limited to relatively low 
redshifts. The PDS galaxies are simply the FIR luminous equivalent of the nearby 
UV-bright SBNGs. The two diagonal lines in Figure 7 suggest that the PDS galaxies 
are flux limited in the FIR at $10^{-10}$ erg cm$^{-2}$ s$^{-1}$ and the MBGs 
at $10^{-11}$ erg cm$^{-2}$ s$^{-1}$. 

\section{Summary and conclusions}
\label{CONCLU}
 
In this contribution, we have confirmed that starburst galaxies 
can be effectively detected and discriminated 
from normal galaxies and AGNs using criteria based on their FIR emission.

In comparison with normal galaxies, the starbursts typically show an excess of
FIR luminosity and distinct colors.
These differences are explained by a higher quantity of hot dust due to the 
higher star formation rates in starbursts as compared to normal galaxies.
In the FIR, the starbursts form a continuous sequence with the normal
galaxies, but can be distinguished based on a color $\alpha$(60,25)$ < -2.5$.  
This color cut--off also marks a variation in the dominant morphological 
types of the galaxies:  normal IRAS galaxies are mostly late-type (Sb and later) spirals,
while starbursts are more numerous among the early-type ones (Sb and earlier). 
This result is consistent with the observations made by
Devereux \& Hameed (1997), and suggests that many early--type spiral galaxies are still
actively forming stars. However, this preference of massive starburst for early--type morphologies
is not new, but a trait of SBNGs (Coziol et al.\ 1997a). 

Like for other samples of SBNGs, no preference for 
barred galaxies is found among the PDS starbursts. 
The role of the bar in starbursts is not well established.
The fact that the number of barred starbursts increases towards
the late--type morphologies may suggest that the bar is important only
in the late--type starbursts. Alternatively, the bar could also have a shorter 
life time in a galaxy with a stronger bulge.

The PDS starbursts are mostly massive (M$_{\rm B} = -20.0 \pm 1.0$). 
This result is consistent with the fact that the 
small-mass \ion{H}{2} galaxies are relatively 
deficient in dust. In the \ion{H}{2} galaxies the dust
may have been swept away 
by starburst winds, or these galaxies may be too young to
have produced enough dust. The FIR colors of the \ion{H}{2} galaxies suggest 
that their dust temperatures are higher than those of the SBNGs. This phenomenon
probably has something to do with the low metallicities of these galaxies, but also
suggests a larger number of young stars than in the SBNGs and, consequently, younger age
for the bursts in the \ion{H}{2} galaxies than in the SBNGs.  

In general, there seems to be no difference between the FIR characteristics
of the UV-bright starbursts and those selected in the FIR.  
The PDS starbursts simply correspond to the FIR luminous branch of the UV-bright 
SBNGs with a mean FIR luminosity Log(L$_{\rm IR}/{\rm L}_\odot) = 10.3 \pm 0.5$
and redshift z $<0.1$.     
The PDS starbursts, in particular, are limited in flux in the FIR at $10^{-10}$ erg 
cm$^{-2}$ s$^{-1}$ while the UV-bright starbursts are limited in flux at $10^{-11}$ 
erg cm$^{-2}$ s$^{-1}$.   

Another interesting result of our analysis is the very few starbursts 
detected in X-rays. Only 4\% of the 200 PDS starbursts were detected by ROSAT. Furthermore, 
the nature of 3 of these X-ray starbursts for which we obtained a spectrum
was found to be ambiguous, the galaxies showing spectral characteristics 
intermediate between those
of starbursts and LINERs. This result, and the fact that
a very high number of Sy1 (71\%), but few Sy2 (22\%) were  
detected in X-rays, cautions against the utilization of the FIR, without other
means of discrimination, to select starburst galaxies in order to
study their X-ray properties. Our observations
suggest, instead, that the contribution by starbursts to the cosmic X-ray background, for example, 
could be negligible (Hasinger 1998).

The relatively high fraction (38\%) of AGNs in our sample is consistent
with observations which suggest that the probability of finding
an AGN increases with the FIR luminosity (de Grijp et al.\ 1987; Veilleux et al.\ 1995).
In our sample, 62\% of the AGNs are Sy2, 34\% are Sy1 and the remaining 4\% are LINERs. 
The higher number of Sy2 encountered is consistent with the idea that these
galaxies are slightly richer in dust than the Sy1 (Malkan et al.\ 1998).
The low fraction of LINERs, on the other hand, is consistent
with the idea that these galaxies are low luminosity AGNs which are not
in a starburst phase (Coziol 1996). 
Only 10\% of the PDS Sy1 and 38\% of the PDS Sy2 have a spectral index in 
the range $-2.5 \leq \alpha(60,25) \leq - 1.9$. But the most striking result of our analysis 
is the fact that only 1\% of the PDS starbursts have a spectral index $\alpha(60,25) > -1.9$. 
The few starbursts which passed this limit are at the extreme of the sequence traced by the
the starburst galaxies and show spectral characteristics at the borderline
between those of starbursts and LINERs. 

Our analysis clearly shows that Seyfert galaxies have 
distinct FIR colors from 
starburst galaxies. This means that the active nucleus in 
AGNs must contribute significantly to the FIR excess emission observed in 
these objects. Now, because this excess is, on average, barely equal to that 
observed in SBNGs, it suggests that the level of star formation in 
Seyfert galaxies may be different from that in starbursts. Taken at face 
value, our results imply that only a small fraction of the Seyfert 
galaxies, maybe $\sim$ 40\% of the Sy2 and $\sim$ 10\% of the Sy1, 
could be dominated by star formation. It is remarkable to find these fractions 
roughly consistent with those determined by Gonz\'alez Delgado et al.\ (1997) based
on their discovery of different star formation properties between 
Sy2 and Sy1 (for similar 
results see also Glass \& Moorwood 1985; Maiolino et al.\ 1995; Hunt et al.\ 
1997 and more recently Malkan et al.\ 1998). A better determination of the
contribution of the active nucleus to the FIR is necessary in order to understand
the relation of AGNs with starbursts.   

\acknowledgments

The electronic version of the tables for the different samples of galaxies defined in this paper (including the
187 IRAS normal galaxies and the Seyfert galaxies) are available upon request to the first author. 
We thank an anonymous referee for useful comments which allow us to improve
the quality of our paper. 
R. C. acknowledges the CNPq for research fellowship. This research has made 
use of the NASA/IPAC Extragalactic Database (NED) which is operated by the Jet 
Propulsion Laboratory, Caltech, under contract with the National Aeronautics 
and Space Administration.

\clearpage

\figcaption[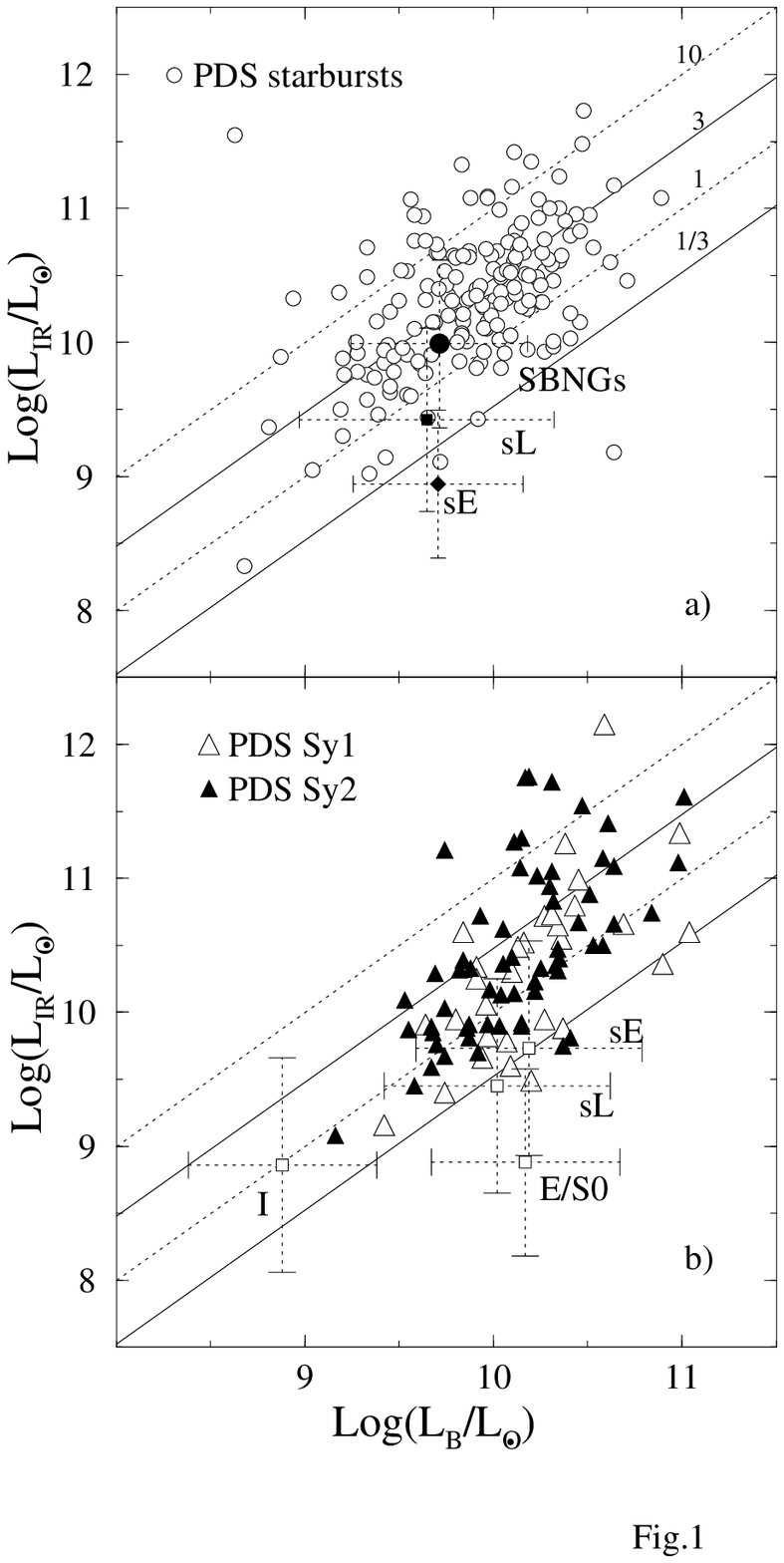]{The FIR vs. B luminosities of a) the PDS starburst candidates, b) the 
PDS Seyfert galaxies. The diagonal lines correspond to the ratios 
L$_{\rm IR}$/L$_{\rm B} =$ 1/3, 1, 3 and 10.
In a), the mean values and dispersions of the luminosities for a sample of
normal late-type 
spirals (sL), early-type spirals (sE) and SBNGs from the literature are indicated.
In b), the mean values and dispersions of the luminosities 
for another sample of normal galaxies (Roberts \& Haynes 1994), which is not biased 
towards emission--line galaxies are also indicated.}

\figcaption[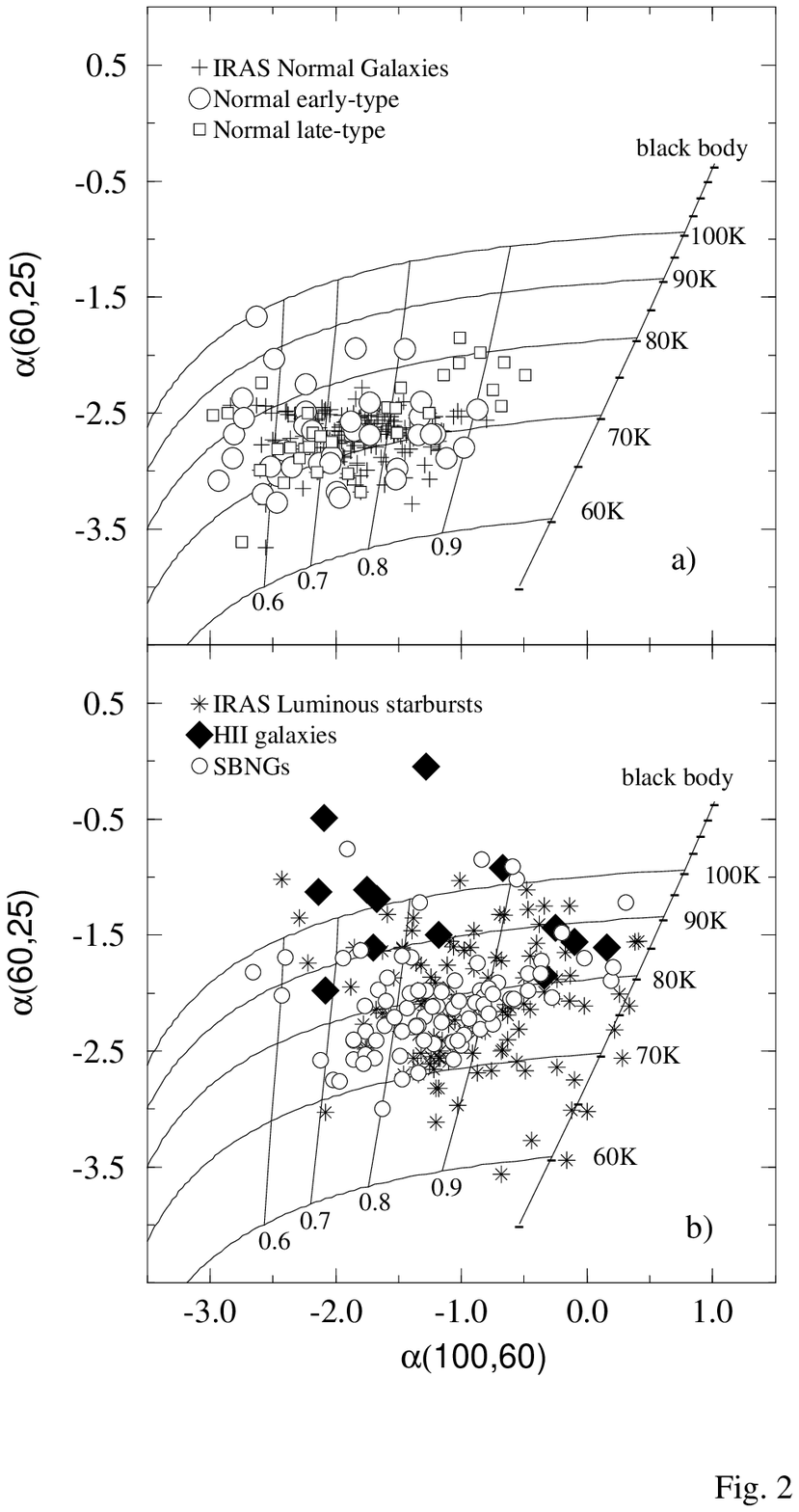]{IRAS color--color diagram for the normal spirals 
and for different samples of starburst galaxies: the SBNGs, \ion{H}{2} galaxies and 
LIGs. The grid corresponds to a two 
blackbody model, composed of a cold component at $\sim 27$K 
added to a hot component, associated to a burst of star formation (Sekiguchi 1987).
The numbers on the grid indicate the fractional contribution of the hot
component to the total FIR luminosity. In a), we call the IRAS Normal galaxies the 
galaxies with a IPSC color $\alpha(60,25) < -2.5$ (see section 3.3).}

\figcaption[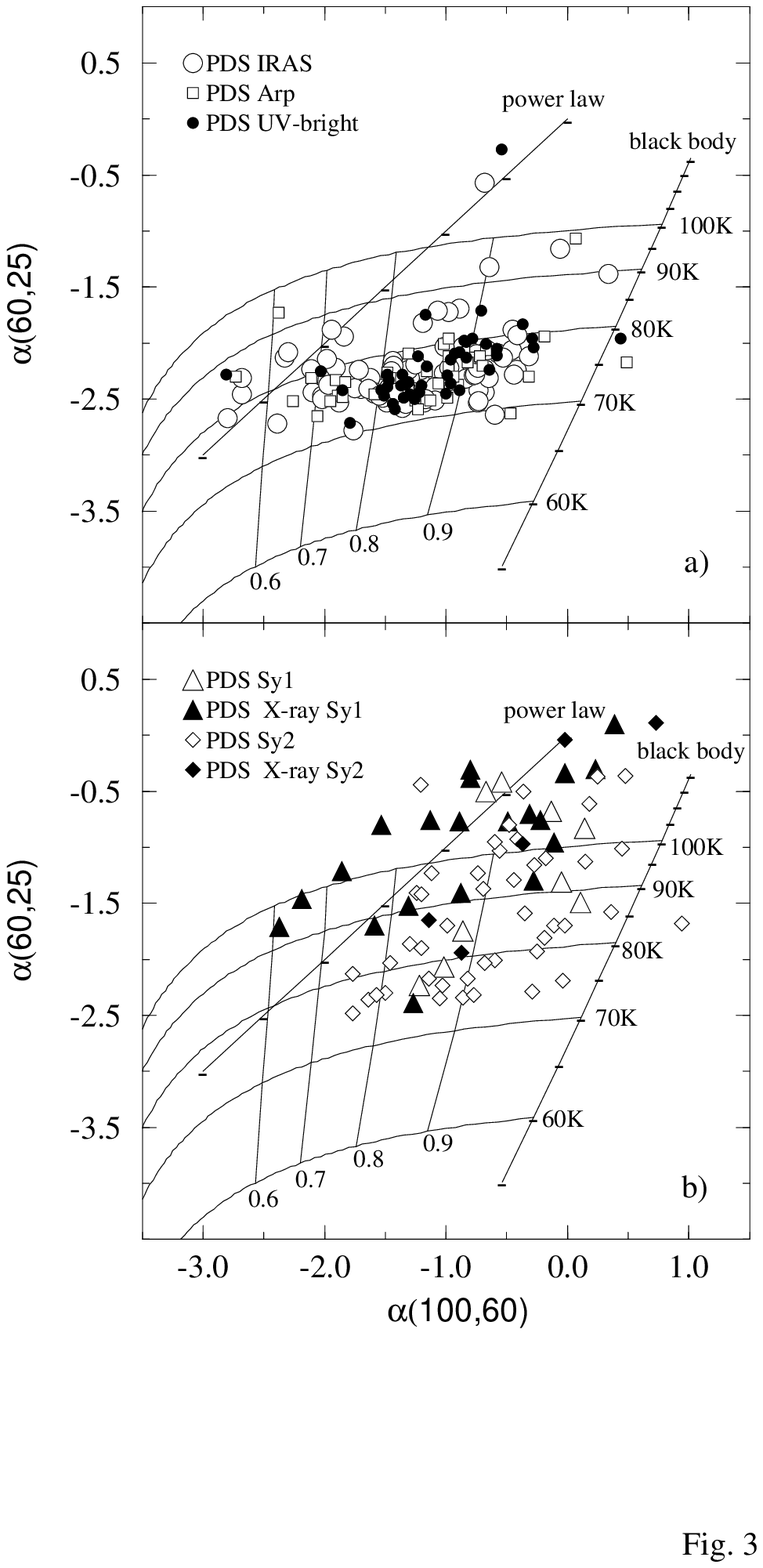]{The same diagram as in Figure 2 for a) the PDS starbursts, b) the PDS Seyfert
galaxies. The locus of a pure power law is also indicated for reference.}

\figcaption[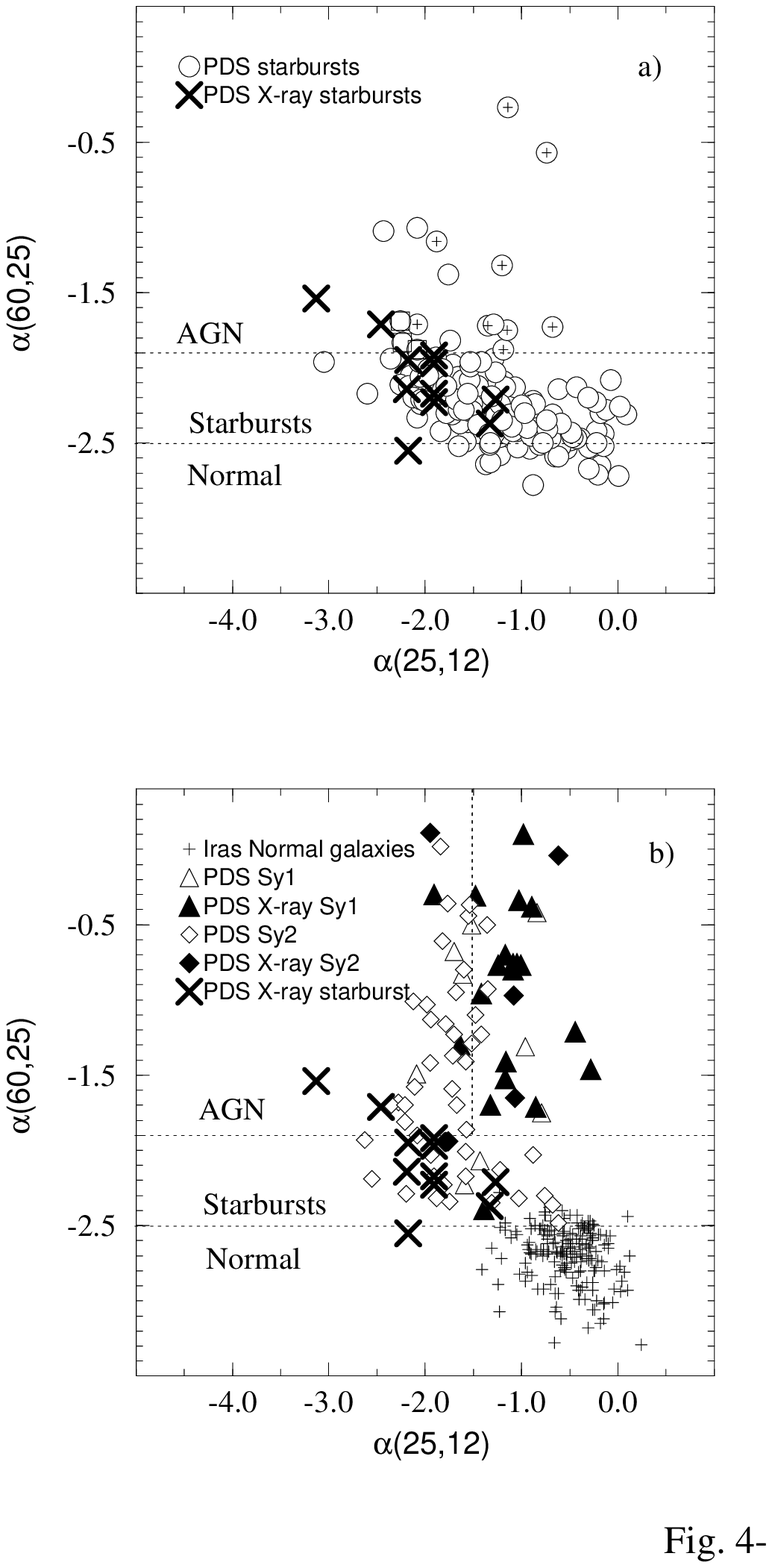]{The color--color diagram $\alpha$(60,25) vs.  
$\alpha$(25,12), showing the distribution of a) the starbursts, b) the AGNs and normal IRAS galaxies.
In a), the misidentified AGNs for which we have spectra are identified by crosses. 
Only 1\% of the starbursts have a spectral index $\alpha$(60,25) $> -1.9$. 
These galaxies (identified by squares) are 
at one of the extreme of the sequence traced by the normal and starburst galaxies.
In b), most of the Sy1 and X-ray AGNs with a color $\alpha$(60,25) $> -1.9$ 
have a color $\alpha$(25,12) $> -1.5$ as opposed to the Sy2 and X-ray starbursts. 
Following our definition in section 3.3, the IRAS normal galaxies have a color $\alpha$(60,25) $< -2.5$.}

\figcaption[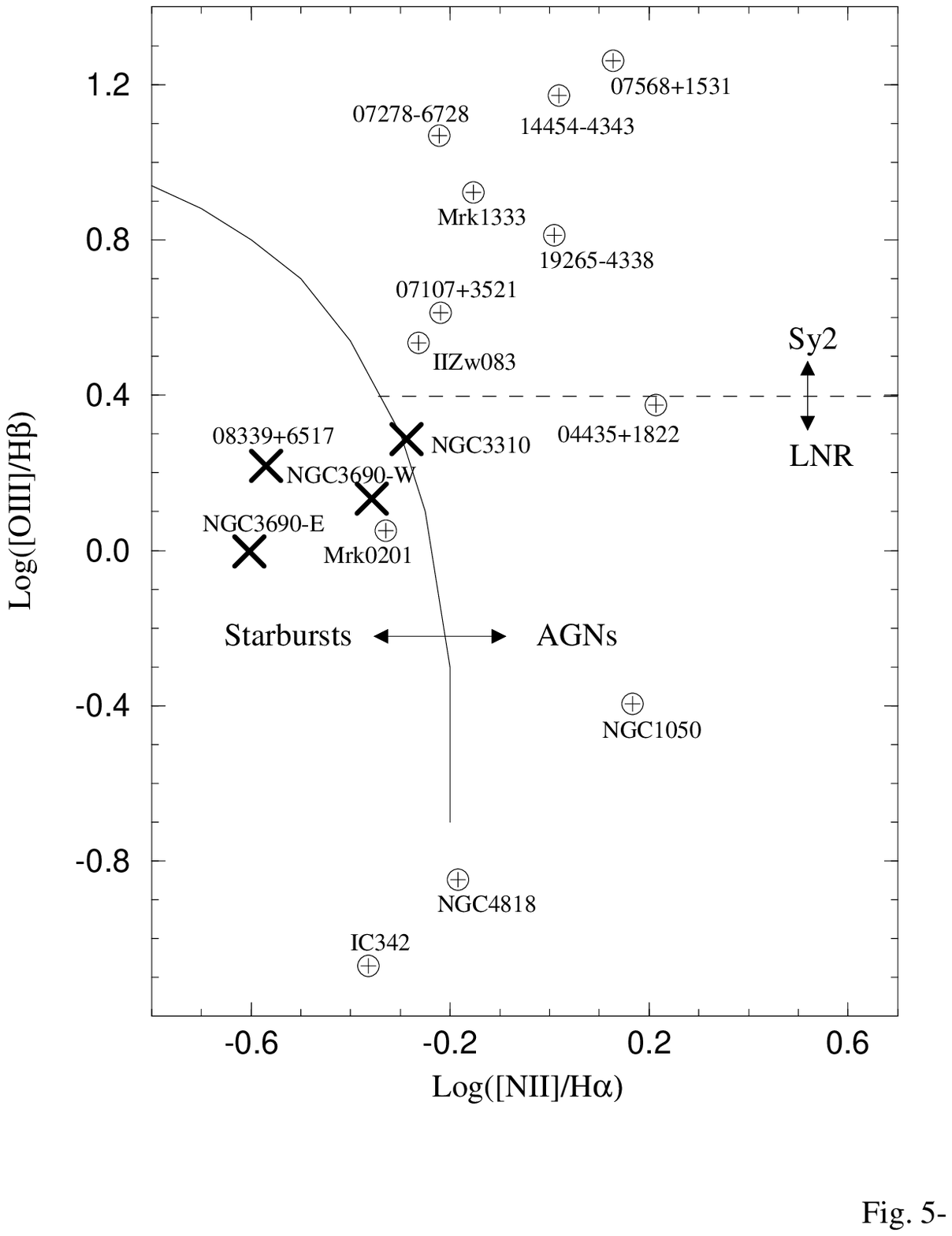]{Diagnostic diagram for classifying some of the PDS galaxies.
The PDS starbursts with a color $\alpha$(60,25) $> -1.9$ are identified by a circle with a cross and the capital
X represent the X-ray PDS starbursts. 9 of the 12 starbursts with a color $\alpha$(60,25) $> -1.9$ turn out
to be misclassified AGNs (mostly Sy2). The X-ray starbursts, on the other hand, have spectral characteristics
suggesting an intermediate nature between one of a LINER and a starburst.}

\figcaption[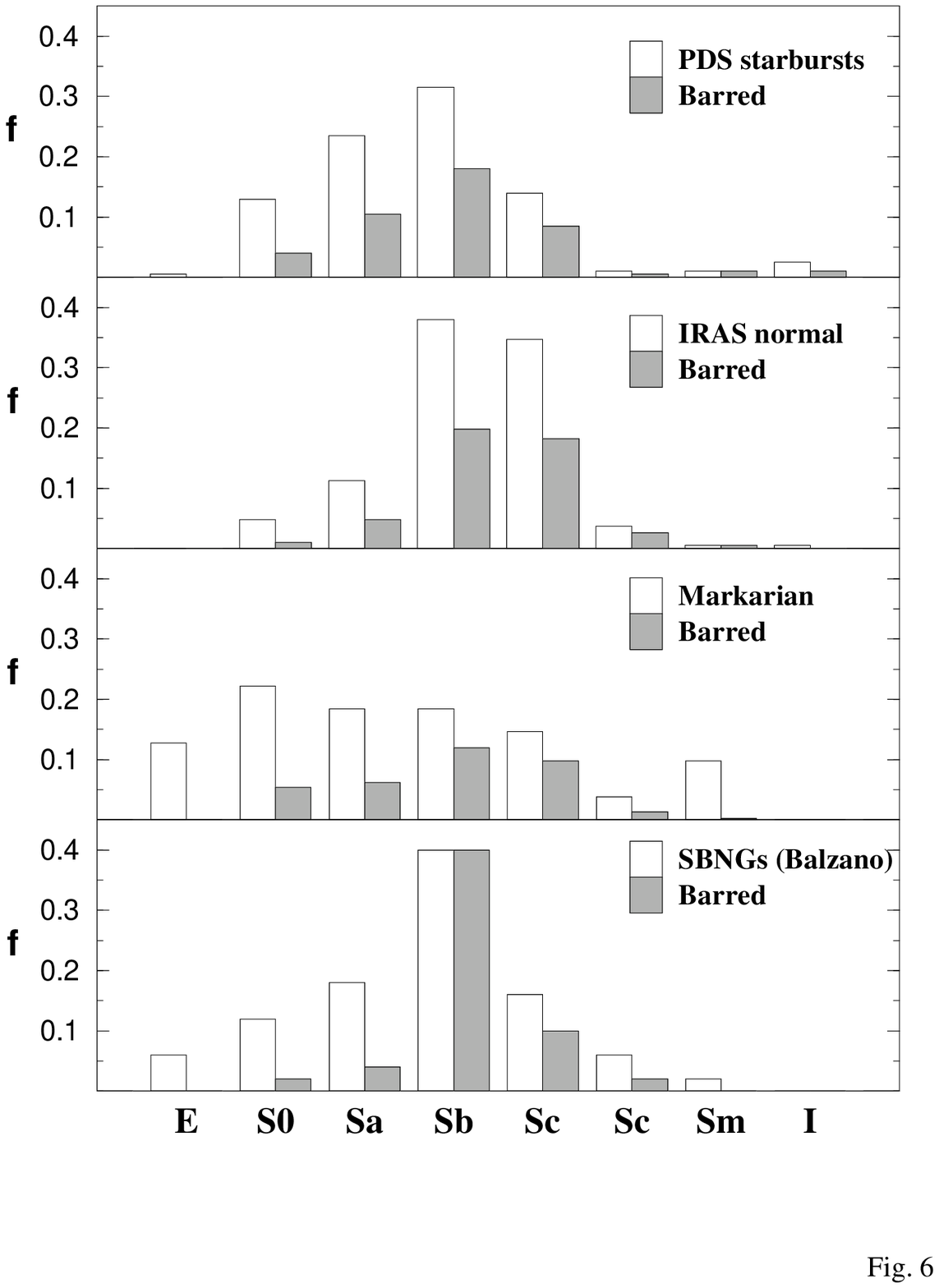]{Distribution of the morphologies for the PDS starbursts, 
the Markarian galaxies and the sample of SBNGs from Balzano (1983). The high number
of early-type PDS and Markarian galaxies (Sb and earlier) is a trait of SBNGs (Coziol et al.\ 1997a).}

\figcaption[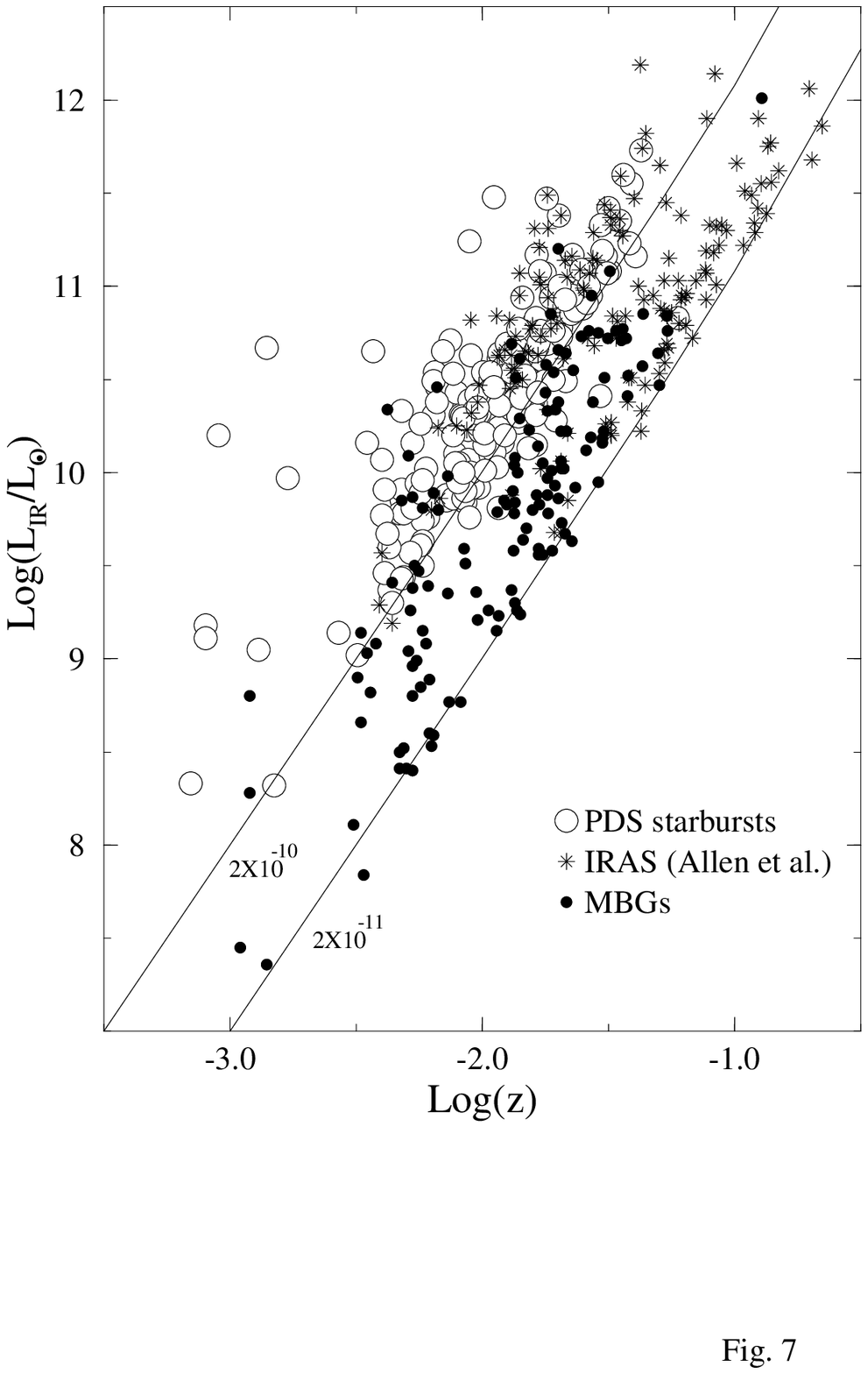]{FIR luminosities vs. redshifts for the PDS starbursts, as compared to
the UV-bright SBNGs (MBG) and a sample of luminous IRAS starbursts  from Allen et al.\ (1991).
The diagonal lines correspond to the flux limits of the different surveys.}

\clearpage
\begin{deluxetable}{llccccc}
\footnotesize
\tablecaption{PDS Starbursts \label{tbl-1}}
\tablewidth{0pt}
\tablehead{
\colhead{IRAS}&\colhead{Other names}&\colhead{$\alpha$}&\colhead{$\delta$} &\colhead{z} &
\colhead{B}   &\colhead{Morph.}\nl
\colhead{ }&\colhead{ }&\colhead{(1950)}&\colhead{(1950)} &\colhead{ } &
\colhead{ } &\colhead{ } 
}
\startdata
 $00013+2028$ & NGC7817                  &  0  1 24.90 &  20 28 18.0 & 0.0083 & 12.56 & SAbc           \nl
 $00022-6220$ & NGC7823                  &  0  2 14.05 & -62 20 23.6 & 0.0148 & 13.40 & SA(s)ab        \nl
 $00073+2538$ & MRK0545, NGC0023         &  0  7 18.40 &  25 38 44.0 & 0.0159 & 12.85 & SB(s)a         \nl
 $00317-2142X$& MBG00317-2142            &  0 31 43.70 & -21 42 51.0 & 0.0268 & 13.66 & (R)SB(rl)bc    \nl
 $00344-3349$ & AM0034-344               &  0 34 25.67 & -33 49 49.0 & 0.0205 &\nodata&\nodata         \nl
 $00345-2945$ & AM0034-294, NGC174       &  0 34 30.99 & -29 45 10.6 & 0.0116 & 13.79 & SB(rs)0/a      \nl
 $00386+4033$ &\nodata                   &  0 38 40.39 &  40 33 40.1 & 0.0116 &\nodata& \nodata        \nl
 $00450-2533X$& NGC0253                  &  0 45 05.75 & -25 33 39.8 & 0.0008 &8.04   &SAB(s)c         \nl
 $00506+7248$ & MCG+12-02-001            &  0 50 40.50 &  72 48 56.0 & 0.0164 &\nodata& \nodata        \nl
 $01053-1746$ & MBG01053-1746, IC1623    &  1  5 18.89 & -17 46 22.1 & 0.0202 &\nodata& \nodata        \nl
 $01076-1707$ & MCG-03-04-014            &  1  7 40.84 & -17  7 11.5 & 0.0351 & 15.00 & \nodata        \nl
 $01171+0308$ & ARP227, NGC470           &  1 17 10.50 &   3  8 53.0 & 0.0083 & 12.53 & SA(rs)b        \nl
 $01384-7515$ & AM0138-751, NGC643B      &  1 38 23.00 & -75 15 54.0 & 0.0127 & 14.57 & SB0            \nl
 $01579+5015$ & UGC01493A                &  1 57 55.29 &  50 15 56.0 & 0.0169 & 14.40 & \nodata        \nl
 $02031-8413$ & AM0203-841               &  2  3  9.00 & -84 13 42.0 & 0.0108 & 13.42 & SAB(rs)c pec   \nl
 $02070-2338$ & MBG02070-2339, AM0207-233&  2  7  0.80 & -23 39  3.4 & 0.0178 & 13.21 & Sbc            \nl
 $02079+3725$ & ARK077, NGC0834          &  2  8  0.02 &  37 25 52.0 & 0.0160 & 13.84 & \nodata        \nl
 $02141-1134$ & MBG02141-1134, NGC873    &  2 14  6.10 & -11 34 48.0 & 0.0136 & 13.00 & Sc pec         \nl
 $02208+4744$ & UGC1845                  &  2 20 51.23 &  47 44 36.6 & 0.0163 & 14.80 & Sab            \nl
 $02315-3915$ & MBG02316-3915, NGC 0986  &  2 31 34.10 & -39 15 49.0 & 0.0067 & 11.64 & (R)SB(rs)b     \nl	
 $02345+2053$ & ARK088, NGC0992, IIZw004 &  2 34 35.60 &  20 53  4.0 & 0.0144 & 13.65 & \nodata        \nl
 $02360-0653$ & NGC1022                  &  2 36  4.30 &  -6 53 24.0 & 0.0053 &\nodata& (R)SB(s)a      \nl
 $02395+3433$ & KUG0239+345, NGC1050     &  2 39 32.06 &  34 33  4.0 & 0.0137 & 13.47 & (R)SB(s)a      \nl
 $03004-2303$ & MBG03004-2303, NGC1187   &  3  0 23.80 & -23  3 47.0 & 0.0048 & 11.34 & SB(r)c         \nl
 $03021+7956$ & UGC02519                 &  3  2 10.50 &  79 56 17.0 & 0.0086 & 14.30 & Scd            \nl
 $03064-0308$ & MRK0603, NGC1222         &  3  6 25.70 &  -3  8 43.0 & 0.0085 & 13.10 & S0- pec        \nl
 $03266+4139$ & NGC1334                  &  3 26 40.66 &  41 39 40.0 & 0.0148 & 14.10 & \nodata        \nl
 $03324-1000$ & MBG03324-1000, NGC1363   &  3 32 25.20 & -10  0 30.0 & 0.0321 & 13.10 & \nodata        \nl
 $03344-2103$ & NGC1377                  &  3 34 25.92 & -21  3 57.3 & 0.0060 & 13.36 & S0             \nl
 $03406+3908$ & MRK1405                  &  3 40 38.17 &  39  8 16.0 & 0.0165 & 13.4  & S0             \nl
 $03419+6756$ & IC0342                   &  3 41 58.60 &  67 56 26.0 & 0.0008 &  9.10 & SAB(rs)cd      \nl
 $03443-1642$ & MCG-03-10-045            &  3 44 20.77 & -16 42 13.6 & 0.0043 & 14.00 & IB pec         \nl
 $03451+6956$ & UGC02866                 &  3 45  7.10 &  69 56 37.0 & 0.0048 & 15.50 & \nodata        \nl
 $03514+1546$ & CGCG 465-012             &  3 51 26.20 &  15 46 55.5 & 0.0227 & 15.20 & Sa             \nl
 $03524-2038$ & NGC1482                  &  3 52 27.11 & -20 38 52.4 & 0.0065 & 13.10 & SA0+ pec       \nl
 $04064+0831$ & NGC1517                  &  4  6 29.00 &   8 31  1.0 & 0.0121 & 14.07 & Scd            \nl
 $04296+2923$ &\nodata                   &  4 29 39.91 &  29 23 39.5 & 0.0075 & 12.19 & \nodata        \nl
 $04326+1904$ & UGC03094                 &  4 32 38.30 &  19  4 14.0 & 0.0253 & 16.50 & \nodata        \nl
 $04389-0257$ & NGC1637                  &  4 38 57.50 &  -2 57 11.0 & 0.0027 & 11.47 & SAB(rs)c       \nl
 $04435+1822$ & UGC03157                 &  4 43 35.74 &  18 22 18.0 & 0.0159 & 15.00 & \nodata        \nl
 $04519+0311$ & MRK1088, NGC1691         &  4 52  1.00 &   3 11 23.0 & 0.0157 & 13.01 & (R)SB(s)0/a    \nl
 $04569-0756$ & NGC1720                  &  4 56 55.60 &  -7 55 59.0 & 0.0144 & 13.00 & SB(s)ab        \nl
 $05053-0805$ & MRK1093, NGC1797         &  5  5 19.50 &  -8  4 59.0 & 0.0151 & 15.50 & (R)SB(rs)a pec \nl
 $05054+1718X$& CGCG468-002              &  5  5 27.39 &  17 18 13.8 & 0.0181 &\nodata&\nodata         \nl
 $05149-3709$ & AM0514-370               &  5 14 55.00 & -37  9 12.0 & 0.0044 & 13.04 & Sbc            \nl
 $06107+7822$ & NGC2146                  &  6 10 42.16 &  78 22 27.6 & 0.0037 & 11.38 & SB(s)ab pec    \nl
 $06141+8220$ & UGC03435                 &  6 14  9.18 &  82 20 31.9 & 0.0150 & 14.72 & \nodata        \nl
 $06189-2001$ & UGCA128                  &  6 18 56.00 & -20  1 24.0 & 0.0067 &\nodata& \nodata        \nl
 $06194-5733$ & AM0619-573, NGC2221      &  6 19 26.00 & -57 33 12.0 & 0.0081 & 13.83 & SA pec         \nl
 $06210+4932$ & CGCG233-017              &  6 21  3.58 &  49 32 13.0 & 0.0202 & 14.90 & \nodata        \nl
 $06259-4708$ & AM0626-470               &  6 26  0.00 & -47  8 36.0 & 0.0392 & 19.26 & \nodata        \nl
 $06399-5828$ & AM0639-582               &  6 39 57.00 & -58 28 36.0 & 0.0086 & 13.10 & (R)SAB(s)bc    \nl
 $07007+8427$ & NGC2268                  &  7  0 52.70 &  84 27 45.0 & 0.0081 & 12.24 & SAB(r)bc       \nl
 $07027-6011$ & AM0702-601               &  7  2 44.50 & -60 11  6.0 & 0.0309 &\nodata& \nodata        \nl
 $07107+3521$ & UGC03752                 &  7 10 45.14 &  35 21 55.0 & 0.0163 & 14.80 & \nodata        \nl
 $07176-3533$ & ESO367-G017              &  7 17 38.00 & -35 33 48.0 & 0.0092 & 13.85 & (R)SB(rl)a     \nl
 $07203+5803$ & UGC03828                 &  7 20 21.50 &  58  4  1.0 & 0.0124 & 12.90 & SAB(rs)b       \nl
 $07204+3332$ & MRK1199, UGC03829        &  7 20 28.30 &  33 32 20.7 & 0.0143 & 13.70 & Sc             \nl
 $07256+3355$ & GC2388                   &  7 25 38.17 &  33 55 17.4 & 0.0144 & 14.67 & \nodata        \nl
 $07278-6728$ & AM0727-672, IC2202       &  7 27 50.00 & -67 28 12.0 & 0.0116 & 13.61 & SAB(s)bc       \nl
 $07369-5504$ & AM0737-550               &  7 36 59.00 & -55  4 30.0 & 0.0091 & 11.85 & SB(s)b         \nl
 $07568+1531$ & UGC04145                 &  7 56 50.30 &  15 31 30.0 & 0.0159 & 14.06 & Sa             \nl
 $08007-6600$ &\nodata                   &  8  0 43.59 & -66  0 51.9 & 0.0407 & 16.18 & pec            \nl
 $08311-2248$ & AM0831-224, NGC2613      &  8 31 11.10 & -22 48  1.0 & 0.0057 & 11.16 & SA(s)b         \nl
 $08339+6517X$&\nodata                   &  8 33 57.30 &  65 17 45.0 & 0.0191 & 14.16 & \nodata        \nl
 $08406-1952$ & ESO563-G014              &  8 40 41.70 & -19 52 19.0 & 0.0058 & 14.00 & SBd            \nl
 $08425+7416$ & ARP080, NGC2633          &  8 42 33.80 &  74 16 54.0 & 0.0079 & 12.90 & SB(s)b         \nl
 $08437-1907$ & NGC2665                  &  8 43 45.00 & -19  7 12.0 & 0.0058 & 13.50 & (R)SB(r)a      \nl
 $09004-2031$ & ESO564-G011              &  9  0 30.00 & -20 31 36.0 & 0.0088 & 14.50 & Sa             \nl
 $09120+4107$ & NGC2785                  &  9 12  3.07 &  41  7 32.6 & 0.0098 & 14.73 & Im             \nl
 $09141+4212$ & ARP283, NGC2798          &  9 14  9.41 &  42 12 34.0 & 0.0064 & 13.04 & SB(s)a pec     \nl
 $09395+0454$ & MRK0708, NGC2966         &  9 39 34.54 &   4 54  6.5 & 0.0072 & 13.56 & \nodata        \nl
 $09399+3204$ & MRK0404, NGC2964         &  9 39 56.90 &  32  4 34.0 & 0.0048 & 12.00 & SAB(r)bc       \nl
 $09434-1408$ & ARP245, NGC2993          &  9 43 24.01 & -14 08 12.6 & 0.0081 & 13.11 & Sa pec         \nl
 $09479+3347$ & KUG0947+337, NGC3021     &  9 47 59.60 &  33 47 18.0 & 0.0058 & 12.91 & SA(rs)bc       \nl
 $09510+0149$ & NGC3044                  &  9 51  6.00 &   1 48 57.0 & 0.0047 & 12.46 & SB(s)c         \nl
 $09511-1214$ & NGC3058                  &  9 51 10.43 & -12 14 45.1 & 0.0250 &\nodata& \nodata        \nl
 $09517+6955X$& M082                     &  9 51 43.48 &  69 55 00.8 & 0.0007 & 9.3   & I0             \nl
 $09578+0336$ & UGC05376                 &  9 57 51.10 &   3 36 56.0 & 0.0072 & 14.02 & Sdm            \nl
 $09593+6858$ & NGC3077                  &  9 59 19.95 &  68 58 29.9 & 0.0007 & 10.61 & I0 pec         \nl
 $10102+1716$ & NGC3154                  & 10 10 18.00 &  17 16 58.0 & 0.0225 &\nodata& \nodata        \nl
 $10138+2122$ & NGC3177                  & 10 13 49.20 &  21 22 28.0 & 0.0049 & 13.04 & SA(rs)b        \nl
 $10153+2205$ & ARP316, NGC3189          & 10 15 20.64 &  22  4 54.9 & 0.0049 & 12.12 & SA(s)a pec     \nl
 $10257-4338X$&AM1025-433, NGC3256       & 10 25 43.40 & -43 38 48.0 & 0.0091 & 12.15 & \nodata        \nl
 $10270-4351$ & AM1027-435, NGC3263      & 10 27  4.50 & -43 51 59.0 & 0.0093 & 12.50 & SB(rs)cd       \nl
 $10292-4148$ & ESO317-G041              & 10 29 12.00 & -41 48 12.0 & 0.0191 & 14.38 & SB(r)bc pec    \nl
 $10356+5345X$& ARP217, NGC3310          & 10 35 40.10 &  53 45 49.0 & 0.0033 & 11.15 & SAB(r)bc pec   \nl
 $10409-4556$ & ESO264-G036              & 10 40 57.00 & -45 56 54.0 & 0.0231 & 14.30 & SB(s)b         \nl
 $10484-0152$ & ARK258, IC0651           & 10 48 25.30 &  -1 53  4.0 & 0.0152 & 13.60 & SB(s)m pec     \nl
 $10560+6147$ & MRK0158, NGC3471         & 10 56  2.20 &  61 47 56.0 & 0.0078 & 13.23 & Sa             \nl
 $10567-4310$ & AM1056-430               & 10 56 44.00 & -43 10 24.0 & 0.0170 & 14.97 & SA(rs)bc       \nl
 $11004+2814$ & KUG1100+282, NGC3504     & 11  0 28.50 &  28 14 31.0 & 0.0057 & 11.67 & (R)SAB(s)ab    \nl
 $11005-1601$ & NGC3508                  & 11  0 30.80 & -16  1  9.0 & 0.0131 & 13.20 & SA(r)b pec     \nl
 $11082-4849$ & ESO215-G031              & 11  8 18.00 & -48 49 54.0 & 0.0088 & 13.64 & (R)SB(r)b      \nl
 $11122-2327$ & AM1112-232, NGC3597      & 11 12 14.20 & -23 27 20.0 & 0.0117 & 13.60 & S0+ pec        \nl
 $11149+0449$ & NGC3611                  & 11 14 54.70 &   4 49 41.0 & 0.0057 & 12.77 & SA(s)a pec     \nl
 $11186-0242$ & CGCG011-076              & 11 18 39.08 &  -2 42 36.5 & 0.0252 & 14.78 & SAB(s)b        \nl
 $11257+5850X$& MRK0171, NGC3690         & 11 25 41.40 &  58 50 15.4 & 0.0104 & 11.8  & SBm pec        \nl
 $11330+7048$ & NGC3735                  & 11 33  4.80 &  70 48 42.0 & 0.0097 & 12.50 & SAc            \nl
 $11442-2738$ & AM1144-273, NGC3885      & 11 44 15.00 & -27 38 42.0 & 0.0060 & 11.89 & SAB(r)0/a      \nl
 $12015+3210$ & KUG1201+321, NGC4062     & 12  1 30.50 &  32 10 26.0 & 0.0032 & 11.90 & SA(s)c         \nl
 $12038+5259$ & NGC4102                  & 12  3 51.66 &  52 59 23.7 & 0.0035 & 11.99 & SAB(s)b        \nl
 $12063+7511$ & NGC4133                  & 12  6 24.80 &  75 10 52.0 & 0.0052 & 13.11 & SABb           \nl
 $12080+1618$ & MRK0759, NGC4152         & 12  8  4.44 &  16 18 40.8 & 0.0077 & 12.66 & SAB(rs)c       \nl
 $12115-4656$ & AM1211-465               & 12 11 34.80 & -46 57  2.0 & 0.0182 & 14.17 & SA(rs)ab       \nl
 $12116+5448X$& MRK0201, NGC4194, IZw033 & 12 11 41.20 &  54 48 15.0 & 0.0090 & 13.01 & IBm pec        \nl
 $12121-3513$ & AM1212-351               & 12 12  8.00 & -35 13 54.0 & 0.0089 & 12.96 & SB(rs)b        \nl
 $12142-4841$ &\nodata                   & 12 14 12.79 & -48 41 39.2 & 0.0177 & 16.20 & E              \nl
 $12173+0537$ & NGC4273                  & 12 17 22.74 &   5 37 13.0 & 0.0083 & 12.39 & SB(s)c         \nl
 $12190+1452$ & NGC4298                  & 12 19  0.55 &  14 53  1.1 & 0.0043 & 12.04 & SA(rs)c        \nl
 $12193-4303$ & AM1219-430               & 12 19 18.00 & -43  3 24.0 & 0.0234 &\nodata& \nodata        \nl
 $12204+6607$ & NGC4332                  & 12 20 27.60 &  66  7 15.0 & 0.0102 & 13.12 & SB(s)a         \nl
 $12221+3939$ & MRK0439, NGC4369         & 12 22  8.03 &  39 39 35.0 & 0.0041 & 12.33 & (R)SA(rs)a     \nl
 $12290+5814$ & MRK0213, NGC4500         & 12 29  2.60 &  58 14 26.0 & 0.0112 & 13.10 & SB(s)a         \nl
 $12319+0227$ & NGC4536                  & 12 31 53.81 &   2 27 50.5 & 0.0064 & 11.16 & SAB(rs)bc      \nl
 $12351-4015$ & NGC4575                  & 12 35  9.00 & -40 15 42.0 & 0.0098 & 13.12 & SAB(rs)bc      \nl
 $12398-0641$ & MRK1333, NGC4628         & 12 39 50.14 &  -6 41 50.4 & 0.0097 & 14.50 & SA(s)b         \nl
 $12456-0303$ & NGC4691                  & 12 45 39.02 &  -3  3 36.9 & 0.0040 & 11.66 & (R)SB(s)0/a pec\nl
 $12498-3845$ & MCG-06-28-02             & 12 49 53.00 & -38 45 24.0 & 0.0141 & 13.29 & (R)SA(rs)b     \nl
 $12542-0815$ & NGC4818                  & 12 54 12.70 &  -8 15 13.0 & 0.0041 & 12.00 & SAB(rs)ab pec  \nl
 $12596-1529$ & MCG-02-33-098            & 12 59 40.80 & -15 29 58.0 & 0.0160 & 14.50 & Sc pec         \nl
 $13063-1515$ & NGC4984                  & 13  6 18.20 & -15 15  1.0 & 0.0042 & 12.25 & (R)SAB(rs)0+   \nl
 $13078-4117$ & ESO323-G090              & 13  7 47.00 & -41 17 36.0 & 0.0102 & 13.92 & SB(rs)0+       \nl
 $13109-4912$ & ESO219-G041              & 13 11  0.00 & -49 12 54.0 & 0.0115 & 12.90 & (R)SB(s)ab     \nl
 $13136+6223$ & ARP238                   & 13 13 42.14 &  62 23 16.2 & 0.0317 & 15.00 & Sc             \nl
 $13166-1434$ & NGC5073                  & 13 16 41.20 & -14 34 50.0 & 0.0093 & 13.00 & SB(s)c         \nl
 $13286-3432$ & AM1328-343, NGC5188      & 13 28 37.00 & -34 32 18.0 & 0.0077 & 12.96 & (R)SAB(rs)b    \nl
 $13301-2356$ & AM1330-235, IC4280       & 13 30  7.90 & -23 57  1.0 & 0.0165 & 13.46 & \nodata       \nl
 $13304+6301$ & ARP104, NGC5218          & 13 30 27.80 &  63  1 27.0 & 0.0102 & 13.10 & SB(s)b pec    \nl
 $13341-2936X$& M083                     & 13 34 11.55 & -29 36 42.2 & 0.0017 &  8.20 & SAB(s)c       \nl
 $13370-3123$ & AM1337-312, NGC5253      & 13 37  5.12 & -31 23 13.2 & 0.0013 & 10.87 & Im pec        \nl
 $13373+0105$ & ARP240, NGC5258          & 13 37 24.60 &   1  5  6.0 & 0.0228 & 12.93 & SA(s)b pec    \nl
 $13478-4848$ & ESO221-IG010             & 13 47 48.00 & -48 48 30.0 & 0.0101 &\nodata& \nodata       \nl
 $13549+4205$ & MRK0281, NGC5383         & 13 54 59.99 &  42  5 22.3 & 0.0081 & 12.05 & (R)SB(rs)b pec\nl
 $13591+5934$ & MRK0799, NGC5430         & 13 59  8.50 &  59 34 16.0 & 0.0108 & 12.72 & SB(s)b        \nl
 $14179-4604$ & IC4402                   & 14 18  0.00 & -46  4 12.0 & 0.0053 & 12.00 & Sb            \nl
 $14187+7149$ & MRK0286, NGC5607, VIIZw547 & 14 18 49.60 &  71 49  3.0 & 0.0260 & 13.90 & Pec         \nl
 $14280+3126$ & NGC5653                  & 14 28  1.00 &  31 26 10.0 & 0.0125 & 12.86 & (R)SA(rs)b    \nl
 $14299+0818$ & ARP049, NGC5665          & 14 29 57.70 &   8 17 58.0 & 0.0078 & 12.66 & SAB(rs)c pec  \nl
 $14376-0004$ & NGC5713                  & 14 37 37.56 & -0  04 29.0 & 0.0063 & 11.84 & SAB(rs)bc pec \nl
 $14430-3728$ & ESO386-G019              & 14 43  2.80 & -37 28 32.0 & 0.0148 & 13.65 & SA(r)0/a      \nl
 $14454-4343$ & ESO273-IG004             & 14 45 26.00 & -43 43 24.0 & 0.0386 &\nodata& \nodata       \nl
 $14483+0519$ & NGC5765                  & 14 48 20.76 &   5 19 17.7 & 0.0015 &\nodata& \nodata       \nl
 $14544-4255$ & AM1454-425, IC4518A      & 14 54 26.00 & -42 55 48.0 & 0.0160 & 15.00 & Sc pec        \nl
 $14545-1900$ & IC1077                   & 14 54 32.00 & -19  0 54.0 & 0.0116 & 13.44 & SA(s)bc       \nl
 $14556-4148$ & NGC5786                  & 14 55 41.00 & -41 48 48.0 & 0.0098 & 12.00 & (R)SAB(s)bc   \nl
 $15005+8343$ & MRK0839                  & 15  0 32.68 &  83 43 16.2 & 0.0136 & 13.82 & \nodata       \nl
 $15042-3608$ & NGC5843                  & 15  4 19.00 & -36  8 12.0 & 0.0139 & 13.11 & SB(s)b        \nl
 $15153+5535$ & NGC5908                  & 15 15 23.00 &  55 35 37.0 & 0.0117 & 12.79 & SA(s)b        \nl
 $15188-1254$ & NGC5915                  & 15 18 47.50 & -12 54 47.0 & 0.0077 & 12.75 & SB(s)ab pec   \nl
 $15243+4150$ & IZw112, ARP090, NGC5930  & 15 24 20.73 &  41 51  0.1 & 0.0095 & 13.60 & SAB(rs)b pec  \nl
 $15268-7757$ & ESO022-G010              & 15 26 52.00 & -77 57 18.0 & 0.0084 & 13.72 & S0            \nl
 $15276+1309$ & NGC5936                  & 15 27 39.70 &  13  9 40.0 & 0.0138 & 13.11 & SB(rs)b       \nl
 $15281-0239$ & NGC5937                  & 15 28  9.90 &  -2 39 33.0 & 0.0095 & 13.17 & (R)SAB(rs)b pec\nl
 $15347+4341$ & IC4564                   & 15 34 44.60 &  43 40 57.0 & 0.0195 & 14.42 & \nodata       \nl
 $15420-7531$ & AM1542-753, NGC5967      & 15 42  6.00 & -75 31  6.0 & 0.0092 & 12.70 & SAB(rs)c      \nl
 $15437+0234$ & NGC5990                  & 15 43 44.80 &   2 34 11.0 & 0.0131 & 13.30 & (R)Sa pec     \nl
 $15467-2914X$& NGC6000                  & 15 46 44.30 & -29 14 06.0 & 0.0070 & 13.01 & SB(s)bc       \nl
 $15496+4724$ & UGC10070                 & 15 49 40.32 &  47 24 14.0 & 0.0205 & 13.98 & \nodata       \nl
 $16030+2040$ & MRK0297, NGC6052         & 16  3  1.40 &  20 40 37.0 & 0.0162 &\nodata& \nodata       \nl
 $16037+2137$ & NGC6060                  & 16  3 41.60 &  21 37  8.0 & 0.0153 & 13.80 & SAB(rs)c      \nl
 $16104+5235$ & MRK0496, NGC6090, IZw135 & 16 10 24.02 &  52 35  4.1 & 0.0300 &\nodata& \nodata       \nl
 $16180+3753$ & IZw141, NGC6120          & 16 18  1.19 &  37 53 36.0 & 0.0312 & 14.60 & Pec           \nl
 $16284+0411$ & CGCG052-037              & 16 28 27.00 &   4 11 24.0 & 0.0248 & 15.04 & \nodata       \nl
 $16301+1955$ & NGC6181                  & 16 30  9.40 &  19 55 48.0 & 0.0084 & 12.49 & SAB(rs)c      \nl
 $16516-0948$ &\nodata                   & 16 51 39.65 &  -9 48 31.3 & 0.0227 & 16.55 & Sc            \nl
 $17091+0803$ & UGC10743                 & 17  9  6.20 &   8  3 16.0 & 0.0089 & 14.74 & Sa            \nl
 $17138-1017$ &\nodata                   & 17 13 50.10 & -10 17 25.0 & 0.0177 & 16.57 & \nodata       \nl
 $17222-5953$ & AM1722-595               & 17 22 16.00 & -59 53 24.0 & 0.0203 & 13.74 & Sbc           \nl
 $17363+8646$ & MRK1116, VIIZw729        & 17 36 22.70 &  86 46 38.8 & 0.0264 & 14.30 & \nodata       \nl
 $17442-6314$ & IC4664                   & 17 44 13.00 & -63 14 18.0 & 0.0163 & 13.64 & (R)SAB(r)b    \nl
 $17467+0807$ & CGCG 055-018             & 17 46 42.66 &   8  7  1.8 & 0.0211 & 15.5  & \nodata       \nl
 $17530+3447$ & ARK534                   & 17 53  4.42 &  34 47  0.8 & 0.0167 & 14.10 & Sab           \nl
 $18093-5744$ & AM1809-574, IC4687       & 18  9 20.10 & -57 44 20.0 & 0.0170 & 14.31 & Sb pec        \nl
 $18095+1458$ & NGC6574                  & 18  9 34.70 &  14 58  3.0 & 0.0080 & 12.83 & SAB(rs)bc     \nl
 $18097-6006$ & AM1809-600               & 18  9 47.00 & -60  6 24.0 & 0.0102 & 13.55 & (R)SAB(rs)a   \nl
 $18131+6820$ & VIIZw778, ARP081, NGC6621& 18 13  9.01 &  68 20 52.5 & 0.0213 & 14.00 & Sb pec        \nl
 $18262+2242$ & UGC11246                 & 18 26 16.88 &  22 42  9.2 & 0.0143 & 14.90 & Sab           \nl
 $18293-3413$ &\nodata                   & 18 29 21.38 & -34 13 41.5 & 0.0180 &\nodata& \nodata       \nl
 $18329+5950$ & VIIZw812, NGC6670A       & 18 32 57.44 &  59 50 57.1 & 0.0296 & 15.70 & \nodata       \nl
 $18375-4150$ & ESO336-G009              & 18 37 35.00 & -41 50 30.0 & 0.0192 & 14.00 & SB(s)b        \nl
 $18425+6036$ & NGC6701                  & 18 42 35.12 &  60 36  4.0 & 0.0139 & 13.01 & (R)SB(s)a     \nl
 $19070+5051$ & NGC6764                  & 19  7  1.22 &  50 51  8.1 & 0.0087 & 12.56 & SB(s)bc       \nl
 $19265-4338$ &\nodata                   & 19 26 34.51 & -43 38 56.8 & 0.0595 & 15.80 & \nodata       \nl
 $19384-7045$ & NGC6808                  & 19 38 28.00 & -70 45  6.0 & 0.0111 & 12.46 & SA(r)ab pec   \nl
 $19517-1241$ & NGC6835                  & 19 51 46.10 & -12 42  4.0 & 0.0055 & 13.41 & SB(s)a        \nl
 $19582-3833$ & AM1958-383               & 19 58 14.50 & -38 33  9.0 & 0.0166 & 13.54 & SA(rs)ab      \nl
 $20192+6634$ & NGC6911                  & 20 19 12.15 &  66 34  6.6 & 0.0090 & 15.10 & SBb           \nl
 $20243-0226$ & IIZw083                  & 20 24 20.10 &  -2 26 34.6 & 0.0294 & 15.20 & \nodata       \nl
 $20272-4738$ & NGC6918                  & 20 27 15.00 & -47 38 24.0 & 0.0057 & 14.42 & (R)SAB(rs)a   \nl
 $20338+5958X$& ARP029, NGC6946          & 20 33 49.24 &  59 58 49.2 & 0.0002 &  9.61 & SAB(rs)cd     \nl
 $20550+1655$ & IIZw096                  & 20 55  5.30 &  16 56  3.0 & 0.0363 &\nodata& \nodata       \nl
 $20551-4250$ & AM2055-425               & 20 55  8.82 & -42 50 45.1 & 0.0426 & 14.74 & Merger        \nl
 $21008-4347$ & ESO286-G035              & 21  0 52.83 & -43 47 31.1 & 0.0171 & 15.00 & \nodata       \nl
 $21087+6557$ & UGC11689                 & 21  8 45.11 &  65 57 50.9 & 0.0103 & 14.40 & SB(r)b        \nl
 $21171-0859$ & NGC7051                  & 21 17 10.55 &  -8 59 41.4 & 0.0086 & 14.00 & SB(r)a pec    \nl
 $21330-3846$ & AM2133-384               & 21 33  5.65 & -38 46  0.5 & 0.0192 & 15.07 & pec           \nl 
 $21457-8145$ & AM2145-814               & 21 45 48.00 & -81 45 54.0 & 0.0080 & 12.23 & SB(s)c        \nl
 $23179+1657$ & IIIZw102, NGC7625        & 23 17 59.90 &  16 57  6.0 & 0.0058 & 12.83 & SA(rs)a pec   \nl
 $23192-4245$ & AM2319-424, NGC7632      & 23 19 16.55 & -42 45 14.9 & 0.0048 & 12.95 & (RL)SB(l)0+   \nl
 $23256+2315$ & MRK0326, NGC7677         & 23 25 36.50 &  23 15 22.0 & 0.0123 & 13.93 & SAB(r)bc      \nl
 $23336+0152X$& MRK0538, NGC7714         & 23 33 40.58 &   1 52 42.3 & 0.0093 & 13.00 & SB(s)b pec    \nl
 $23568+2028$ & MRK0332, NGC7798         & 23 56 52.00 &  20 28 17.0 & 0.0084 & 12.97 & \nodata       \nl
\enddata
\end{deluxetable} 
\clearpage
\clearpage
\begin{deluxetable}{lccccccc}
\footnotesize
\tablecaption{PDS starbursts Optical--FIR characteristics \label{tbl-2}}
\tablewidth{0pt}
\tablehead{
\colhead{IRAS}&\colhead{M$_{B}$}&\colhead{Log(L$_{B}$/L$_\odot$)} &\colhead{Log(L$_{IR}$/L$_\odot$)} &
\colhead{$\alpha$(25,12)} &\colhead{$\alpha$(60,25)} 
&\colhead{$\alpha$(100,60)} &\colhead{quality} 
}
\startdata
$00013+2028$ & -20.18 &  9.96 & 10.10 & -0.15 & -2.44 & -2.10 & 3332\nl
$00022-6220$ & -20.46 & 10.07 & 10.37 & -0.81 & -2.46 & -1.32 & 3332\nl
$00073+2538$ & -21.30 & 10.41 & 10.80 & -0.99 & -2.45 & -1.00 & 3232\nl
$00317-2142$ & -21.56 & 10.51 & 10.95 & -1.27 & -2.21 & -1.53 & 3332\nl
$00344-3349$ &\nodata &\nodata& 10.78 & -2.44 & -1.09 &  0.50 & 3332\nl
$00345-2945$ & -21.14 & 10.35 & 10.61 & -1.58 & -2.49 & -0.99 & 3332\nl
$00386+4033$ &\nodata &\nodata&\nodata&  0.09 & -2.31 & -2.68 & 2322\nl
$00450-2533$ & -19.83 &  9.82 & 10.20 & -2.19 & -2.15 & -0.46 & 3332\nl
$00506+7248$ &\nodata &\nodata& 11.17 & -2.15 & -2.05 & -0.34 & 3222\nl
$01053-1746$ &\nodata &\nodata& 11.38 & -2.26 & -2.11 & -0.58 & 3332\nl
$01076-1707$ & -20.79 & 10.20 & 11.35 & -1.41 & -2.33 & -0.93 & 3332\nl
$01171+0308$ & -20.15 &  9.95 & 10.11 & -1.18 & -2.09 & -1.31 & 3232\nl
$01384-7515$ & -19.13 &  9.54 & 10.53 & -1.21 & -2.51 & -1.25 & 3332\nl
$01579+5015$ & -20.96 & 10.27 & 10.53 & -1.24 & -2.10 & -0.75 & 3322\nl
$02031-8413$ & -20.20 &  9.97 & 10.15 & -0.71 & -2.47 & -1.89 & 3332\nl
$02070-2338$ & -21.06 & 10.31 & 10.58 & -0.22 & -2.42 & -1.85 & 3332\nl
$02079+3725$ & -20.34 & 10.02 & 10.68 & -0.96 & -2.46 & -1.30 & 3332\nl
$02141-1134$ & -20.70 & 10.17 & 10.51 & -0.73 & -2.37 & -1.33 & 2232\nl
$02208+4744$ & -20.57 & 10.12 & 10.83 & -1.37 & -2.64 & -0.60 & 3322\nl
$02315-3915$ & -22.07 & 10.71 & 10.46 & -1.77 & -2.28 & -1.36 & 3332\nl
$02345+2053$ & -20.56 & 10.11 & 10.76 & -1.06 & -2.42 & -0.89 & 3332\nl
$02360-0653$ &\nodata &\nodata& 10.16 & -2.07 & -2.09 & -0.57 & 3332\nl
$02395+3433$ & -20.42 & 10.06 & 10.45 & -2.08 & -1.71 & -0.71 & 3332\nl
$03004-2303$ & -20.10 &  9.93 &  9.85 & -1.37 & -2.28 & -1.48 & 3332\nl
$03021+7956$ & -19.29 &  9.61 &  9.84 & -0.58 & -2.46 & -1.89 & 3322\nl
$03064-0308$ & -19.68 &  9.76 & 10.35 & -2.03 & -2.04 & -0.28 & 3332\nl
$03266+4139$ & -20.86 & 10.23 & 10.48 & -0.55 & -2.48 & -1.47 & 3322\nl
$03324-1000$ & -22.51 & 10.89 & 11.08 & -0.62 & -2.39 & -1.48 & 3332\nl
$03344-2103$ & -18.61 &  9.33 &  9.76 & -1.92 & -1.58 &  0.46 & 3332\nl
$03406+3908$ & -21.83 & 10.62 & 10.60 & -0.53 & -2.54 & -1.44 & 3322\nl
$03419+6756$ & -21.87 & 10.64 &  9.18 & -2.26 & -1.69 & -0.89 & 3322\nl
$03443-1642$ & -17.31 &  8.81 &  9.37 & -1.60 & -2.34 & -0.98 & 3332\nl
$03451+6956$ & -17.63 &  8.94 & 10.33 & -1.84 & -2.06 & -0.46 & 3322\nl
$03514+1546$ & -20.65 & 10.15 & 10.89 & -1.21 & -2.32 & -0.79 & 3332\nl
$03524-2038$ & -19.05 &  9.51 & 10.54 & -1.47 & -2.23 & -0.68 & 3332\nl
$04064+0831$ & -19.87 &  9.84 & 10.16 & -0.60 & -2.53 & -1.49 & 3322\nl
$04296+2923$ & -21.61 & 10.53 & 10.71 & -2.17 & -2.12 & -0.32 & 3322\nl
$04326+1904$ & -19.99 &  9.88 & 11.08 & -1.01 & -2.36 & -1.34 & 3322\nl
$04389-0257$ & -18.85 &  9.43 &  9.14 & -1.29 & -1.94 & -1.84 & 3332\nl
$04435+1822$ & -20.37 & 10.04 & 10.49 & -1.19 & -1.88 & -1.94 & 3322\nl
$04519+0311$ & -21.19 & 10.36 & 10.65 & -1.56 & -2.36 & -0.96 & 3332\nl
$04569-0756$ & -21.00 & 10.29 & 10.62 & -1.22 & -2.58 & -1.36 & 3322\nl
$05053-0805$ & -18.61 &  9.33 & 10.71 & -1.89 & -2.09 & -0.93 & 3322\nl
$05054+1718$ &\nodata &\nodata& 10.94 & -2.19 & -2.55 & -0.35 & 3322\nl
$05149-3709$ & -18.29 &  9.20 &  9.30 & -0.18 & -2.65 & -2.06 & 3332\nl
$06107+7822$ & -19.76 &  9.79 & 10.65 & -1.41 & -2.29 & -0.67 & 3332\nl
$06141+8220$ & -19.42 &  9.65 & 10.42 & -1.10 & -2.27 & -1.02 & 3332\nl
$06189-2001$ &\nodata &\nodata&  9.85 & -0.81 & -2.33 & -1.68 & 3322\nl
$06194-5733$ & -18.89 &  9.44 &  9.98 & -1.22 & -2.46 & -1.59 & 3332\nl
$06210+4932$ & -20.25 &  9.99 & 10.65 & -0.56 & -2.49 & -1.52 & 3322\nl
$06259-4708$ & -16.85 &  8.63 & 11.55 & -2.05 & -2.06 & -0.67 & 3332\nl
$06399-5828$ & -19.93 &  9.86 & 10.01 & -0.49 & -2.48 & -1.85 & 3332\nl
$07007+8427$ & -20.51 & 10.09 & 10.05 & -0.39 & -2.47 & -2.02 & 3332\nl
$07027-6011$ &\nodata &\nodata& 11.16 & -1.70 & -2.13 & -0.80 & 3332\nl
$07107+3521$ & -19.52 &  9.69 & 10.15 & -1.20 & -1.32 & -0.64 & 3322\nl
$07176-3533$ & -20.14 &  9.95 &  9.93 & -1.11 & -2.36 & -1.20 & 3322\nl
$07203+5803$ & -20.74 & 10.18 & 10.25 & -0.88 & -2.22 & -1.91 & 3332\nl
$07204+3332$ & -20.27 & 10.00 & 10.55 & -2.06 & -1.96 & -0.29 & 3332\nl
$07256+3355$ & -19.36 &  9.63 & 10.94 & -2.08 & -2.33 & -0.77 & 3332\nl
$07278-6728$ & -20.38 & 10.04 &  9.81 & -0.68 & -1.73 & -2.38 & 3332\nl
$07369-5504$ & -21.43 & 10.46 & 10.15 & -0.70 & -2.52 & -1.95 & 3322\nl
$07568+1531$ & -20.05 &  9.91 & 10.39 & -1.35 & -1.72 & -0.98 & 3332\nl
$08007-6600$ & -20.53 & 10.10 & 11.16 & -1.76 & -1.38 &  0.34 & 3322\nl
$08311-2248$ & -20.96 & 10.27 &  9.93 & -0.18 & -2.30 & -2.73 & 3322\nl
$08339+6517$ & -20.47 & 10.08 & 10.75 & -1.91 & -1.92 & -0.19 & 3332\nl
$08406-1952$ & -18.25 &  9.19 &  9.50 &  0.01 & -2.72 & -2.39 & 3322\nl
$08425+7416$ & -19.65 &  9.75 & 10.43 & -1.68 & -2.21 & -0.98 & 3332\nl
$08437-1907$ & -18.70 &  9.37 &  9.74 & -1.70 & -2.03 & -0.86 & 3322\nl
$09004-2031$ & -18.90 &  9.45 & 10.23 & -2.05 & -2.24 & -0.42 & 3322\nl
$09120+4107$ & -18.23 &  9.18 & 10.37 & -1.00 & -2.45 & -1.20 & 3332\nl
$09141+4212$ & -19.02 &  9.50 & 10.31 & -1.86 & -2.08 & -0.85 & 3332\nl
$09395+0454$ & -18.83 &  9.42 &  9.85 & -1.51 & -2.24 & -0.64 & 3332\nl
$09399+3204$ & -19.45 &  9.67 &  9.91 & -0.97 & -2.49 & -1.35 & 3332\nl
$09434-1408$ & -19.73 &  9.78 & 10.31 & -1.78 & -2.19 & -0.90 & 3332\nl
$09479+3347$ & -18.90 &  9.45 &  9.63 & -0.21 & -2.71 & -1.79 & 3332\nl
$09510+0149$ & -18.95 &  9.47 &  9.78 & -0.91 & -2.54 & -1.33 & 3332\nl
$09511-1214$ &\nodata &\nodata& 10.88 & -1.10 & -2.39 & -1.66 & 3332\nl
$09517+6955$ & -19.52 &  9.70 & 10.67 & -2.10 & -1.60 & -0.11 & 3332\nl
$09578+0336$ & -18.29 &  9.20 &  9.88 & -0.95 & -2.41 & -1.53 & 3332\nl
$09593+6858$ & -16.98 &  8.68 &  8.33 & -1.69 & -2.24 & -1.30 & 3332\nl
$10102+1716$ &\nodata &\nodata& 10.85 & -0.30 & -2.53 & -1.18 & 3332\nl
$10138+2122$ & -18.48 &  9.28 &  9.78 & -0.87 & -2.44 & -1.18 & 3332\nl
$10153+2205$ & -19.41 &  9.65 &  9.44 & -0.15 & -2.52 & -2.26 & 3232\nl
$10257-4338$ & -21.17 & 10.35 & 11.24 & -2.18 & -1.95 & -0.52 & 3322\nl
$10270-4351$ & -20.90 & 10.25 & 10.30 & -0.81 & -2.33 & -1.91 & 3322\nl
$10292-4148$ & -20.56 & 10.11 & 10.61 & -1.08 & -2.16 & -1.44 & 3322\nl
$10356+5345$ & -19.85 &  9.83 & 10.07 & -1.90 & -2.17 & -0.49 & 3332\nl
$10409-4556$ & -21.38 & 10.44 & 10.96 & -0.76 & -2.41 & -1.70 & 3322\nl
$10484-0152$ & -20.39 & 10.04 & 10.51 & -0.90 & -2.42 & -1.53 & 3332\nl
$10560+6147$ & -19.24 &  9.58 & 10.10 & -1.82 & -2.15 & -0.96 & 3332\nl
$10567-4310$ & -19.96 &  9.87 & 10.68 & -1.14 & -2.25 & -1.16 & 3322\nl
$11004+2814$ & -20.15 &  9.95 & 10.26 & -1.71 & -1.98 & -0.85 & 3332\nl
$11005-1601$ & -20.53 & 10.10 & 10.54 & -1.06 & -2.31 & -1.43 & 3232\nl
$11082-4849$ & -19.86 &  9.83 & 10.07 & -1.00 & -2.39 & -1.09 & 3322\nl
$11122-2327$ & -19.96 &  9.87 & 10.64 & -1.52 & -2.12 & -0.72 & 3332\nl
$11149+0449$ & -19.14 &  9.54 &  9.61 & -1.05 & -2.22 & -0.97 & 3232\nl
$11186-0242$ & -20.36 & 10.03 & 10.99 & -1.09 & -2.24 & -1.20 & 3322\nl
$11257+5850$ & -21.44 & 10.47 & 11.48 & -2.46 & -1.71 & -0.07 & 3332\nl
$11330+7048$ & -20.45 & 10.07 & 10.33 & -0.62 & -2.14 & -1.98 & 3332\nl
$11442-2738$ & -20.36 & 10.03 & 10.02 & -1.32 & -2.63 & -0.47 & 3232\nl
$12015+3210$ & -18.62 &  9.34 &  9.02 & -0.12 & -2.28 & -2.81 & 3332\nl
$12038+5259$ & -18.74 &  9.38 & 10.16 & -2.04 & -2.21 & -0.78 & 3332\nl
$12063+7511$ & -18.61 &  9.33 &  9.57 & -1.32 & -2.36 & -1.43 & 3332\nl
$12080+1618$ & -19.83 &  9.82 &  9.86 & -0.80 & -2.33 & -1.47 & 3232\nl
$12115-4656$ & -20.60 & 10.13 & 10.69 & -0.98 & -2.36 & -1.06 & 3322\nl
$12116+5448$ & -19.78 &  9.80 & 10.63 & -2.24 & -1.83 & -0.37 & 3332\nl
$12121-3513$ & -20.09 &  9.92 & 10.39 & -2.12 & -2.20 & -0.70 & 3332\nl
$12142-4841$ & -18.61 &  9.33 & 10.49 & -0.43 & -2.47 & -1.55 & 3322\nl
$12173+0537$ & -20.22 &  9.98 & 10.30 & -0.83 & -2.31 & -1.62 & 2232\nl
$12190+1452$ & -19.18 &  9.56 &  9.60 & -0.30 & -2.67 & -2.80 & 3332\nl
$12193-4303$ &\nodata &\nodata& 10.87 & -0.81 & -2.51 & -1.13 & 3222\nl
$12204+6607$ & -19.93 &  9.86 & 10.31 & -1.61 & -2.31 & -1.30 & 3332\nl
$12221+3939$ & -18.76 &  9.39 &  9.46 & -1.15 & -2.44 & -1.22 & 3332\nl
$12290+5814$ & -20.15 &  9.95 & 10.11 & -1.07 & -2.13 & -0.83 & 3332\nl
$12319+0227$ & -20.87 & 10.23 & 10.49 & -1.28 & -2.44 & -0.68 & 3332\nl
$12351-4015$ & -20.31 & 10.01 & 10.32 & -0.84 & -2.44 & -1.59 & 3332\nl
$12398-0641$ & -18.49 &  9.28 &  9.92 & -1.15 & -1.75 & -1.17 & 3232\nl
$12456-0303$ & -19.37 &  9.64 &  9.77 & -1.66 & -2.06 & -0.66 & 3332\nl
$12498-3845$ & -20.66 & 10.15 & 10.27 & -0.15 & -2.49 & -1.93 & 3332\nl
$12542-0815$ & -19.14 &  9.54 &  9.91 & -2.08 & -1.88 & -0.45 & 3332\nl
$12596-1529$ & -19.57 &  9.71 & 10.67 & -1.89 & -1.93 & -0.41 & 2332\nl
$13063-1515$ & -18.92 &  9.45 &  9.67 & -1.17 & -2.13 & -0.53 & 3332\nl
$13078-4117$ & -19.49 &  9.68 & 10.15 & -1.04 & -2.53 & -0.74 & 3332\nl
$13109-4912$ & -21.32 & 10.41 & 10.03 & -0.43 & -2.13 & -2.33 & 3322\nl
$13136+6223$ & -20.55 & 10.11 & 11.42 & -2.36 & -1.94 & -0.19 & 3332\nl
$13166-1434$ & -19.96 &  9.87 & 10.33 & -1.84 & -2.42 & -0.73 & 3232\nl
$13286-3432$ & -19.63 &  9.74 & 10.53 & -1.64 & -2.37 & -0.90 & 3332\nl
$13301-2356$ & -20.92 & 10.26 & 10.67 & -0.65 & -2.59 & -1.23 & 3332\nl
$13304+6301$ & -19.99 &  9.88 & 10.33 & -1.72 & -2.31 & -1.40 & 3332\nl
$13341-2936$ & -21.08 & 10.32 &  9.97 & -1.91 & -1.97 & -1.49 & 3332\nl
$13370-3123$ & -17.88 &  9.04 &  9.05 & -2.08 & -1.07 &  0.07 & 3332\nl
$13373+0105$ & -21.88 & 10.64 & 11.17 & -1.06 & -2.31 & -1.46 & 3332\nl
$13478-4848$ &\nodata &\nodata& 10.54 & -1.26 & -2.32 & -0.84 & 3322\nl
$13549+4205$ & -20.51 & 10.09 & 10.05 & -0.91 & -2.25 & -2.03 & 3232\nl
$13591+5934$ & -20.46 & 10.07 & 10.54 & -1.45 & -2.12 & -1.23 & 3332\nl
$14179-4604$ & -20.06 &  9.91 &  9.81 & -1.22 & -2.24 & -1.31 & 3322\nl
$14187+7149$ & -21.23 & 10.38 & 10.91 & -1.61 & -2.08 & -0.89 & 3332\nl
$14280+3126$ & -20.63 & 10.14 & 10.69 & -0.82 & -2.44 & -1.26 & 3332\nl
$14299+0818$ & -19.86 &  9.83 & 10.05 & -1.09 & -2.42 & -1.31 & 3332\nl
$14376-0004$ & -20.37 & 10.04 & 10.38 & -1.17 & -2.33 & -1.18 & 3332\nl
$14430-3728$ & -20.51 & 10.09 & 10.52 & -1.05 & -2.42 & -0.73 & 3322\nl
$14454-4343$ &\nodata &\nodata& 11.23 & -1.88 & -1.16 & -0.06 & 3322\nl
$14483+0519$ &\nodata &\nodata&  8.32 & -1.29 & -1.71 & -1.07 & 3332\nl
$14544-4255$ & -19.54 &  9.70 & 10.73 & -1.82 & -2.01 & -1.02 & 3322\nl
$14545-1900$ & -20.22 &  9.98 & 10.24 & -0.84 & -2.40 & -1.33 & 3332\nl
$14556-4148$ & -21.31 & 10.41 & 10.22 & -0.73 & -2.40 & -1.75 & 3322\nl
$15005+8343$ & -20.05 &  9.91 & 10.40 & -1.53 & -1.99 & -0.83 & 3332\nl
$15042-3608$ & -20.92 & 10.26 & 10.30 & -0.60 & -2.53 & -1.88 & 3322\nl
$15153+5535$ & -20.57 & 10.12 & 10.36 & -0.47 & -2.46 & -2.68 & 3332\nl
$15188-1254$ & -20.25 &  9.99 & 10.20 & -1.54 & -2.29 & -0.77 & 3332\nl
$15243+4150$ & -19.37 &  9.64 & 10.32 & -1.97 & -2.01 & -0.67 & 3332\nl
$15268-7757$ & -19.28 &  9.60 &  9.86 & -1.54 & -2.08 & -0.79 & 3322\nl
$15276+1309$ & -20.69 & 10.16 & 10.67 & -1.31 & -2.19 & -1.25 & 3332\nl
$15281-0239$ & -20.11 &  9.93 & 10.42 & -0.78 & -2.45 & -1.44 & 3332\nl
$15347+4341$ & -20.13 &  9.94 & 10.28 & -0.48 & -2.43 & -1.96 & 3332\nl
$15420-7531$ & -20.42 & 10.06 &  9.92 & -0.68 & -2.31 & -2.11 & 3322\nl
$15437+0234$ & -20.59 & 10.12 & 10.64 & -1.27 & -2.03 & -1.01 & 3332\nl
$15467-2914$ & -19.89 &  9.84 & 10.65 & -1.91 & -2.23 & -0.77 & 3322\nl
$15496+4724$ & -20.64 & 10.14 & 10.73 & -0.22 & -2.50 & -2.02 & 3332\nl
$16030+2040$ &\nodata &\nodata& 10.65 & -1.51 & -2.29 & -0.99 & 3332\nl
$16037+2137$ & -20.32 & 10.02 & 10.13 & -0.07 & -2.08 & -2.30 & 3332\nl
$16104+5235$ &\nodata &\nodata& 11.19 & -1.97 & -2.05 & -0.58 & 3332\nl
$16180+3753$ & -20.88 & 10.24 & 11.07 & -0.98 & -2.38 & -1.37 & 3332\nl
$16284+0411$ & -20.20 &  9.97 & 11.09 & -1.33 & -2.51 & -1.09 & 3332\nl
$16301+1955$ & -20.37 & 10.04 & 10.29 & -0.90 & -2.45 & -1.59 & 3332\nl
$16516-0948$ & -19.24 &  9.58 & 10.95 & -0.88 & -2.78 & -1.76 & 3322\nl
$17091+0803$ & -18.31 &  9.21 &  9.76 & -1.12 & -2.31 & -0.65 & 3332\nl
$17138-1017$ & -19.17 &  9.56 & 11.07 & -1.65 & -2.28 & -0.44 & 3322\nl
$17222-5953$ & -21.15 & 10.35 & 11.00 & -1.78 & -2.30 & -0.32 & 3322\nl
$17363+8646$ & -21.03 & 10.30 & 11.00 & -0.62 & -2.59 & -1.42 & 3332\nl
$17442-6314$ & -20.76 & 10.19 & 10.31 & -0.23 & -2.23 & -2.11 & 3322\nl
$17467+0807$ & -19.79 &  9.80 & 10.49 & -1.40 & -1.23 &  0.05 & 3322\nl
$17530+3447$ & -20.19 &  9.96 & 10.70 & -0.70 & -2.47 & -1.51 & 3332\nl
$18093-5744$ & -20.20 &  9.97 & 11.08 & -1.78 & -2.12 & -0.98 & 3322\nl
$18095+1458$ & -20.53 & 10.10 & 10.41 & -0.94 & -2.40 & -1.47 & 3322\nl
$18097-6006$ & -19.89 &  9.84 & 10.21 & -1.42 & -1.96 & -0.98 & 3322\nl
$18131+6820$ & -20.87 & 10.24 & 10.93 & -1.68 & -2.21 & -1.16 & 3322\nl
$18262+2242$ & -19.56 &  9.71 & 10.40 & -1.10 & -2.39 & -1.36 & 3322\nl
$18293-3413$ &\nodata &\nodata& 11.47 & -1.65 & -2.52 & -0.73 & 2322\nl
$18329+5950$ & -19.86 &  9.83 & 11.33 & -1.45 & -2.38 & -1.20 & 3332\nl
$18375-4150$ & -20.76 & 10.19 & 10.50 & -0.31 & -2.20 & -1.43 & 3322\nl
$18425+6036$ & -20.95 & 10.27 & 10.77 & -1.34 & -2.44 & -1.40 & 3332\nl
$19070+5051$ & -20.39 & 10.04 & 10.13 & -1.74 & -1.82 & -1.19 & 3322\nl
$19265-4338$ & -21.42 & 10.46 & 10.83 & -0.74 & -0.57 & -0.68 & 3332\nl
$19384-7045$ & -21.07 & 10.32 & 10.46 & -0.59 & -2.37 & -1.97 & 3332\nl
$19517-1241$ & -18.84 &  9.42 &  9.94 & -1.60 & -2.28 & -0.76 & 3322\nl
$19582-3833$ & -20.90 & 10.25 & 10.43 &  0.02 & -2.26 & -1.48 & 3332\nl
$20192+6634$ & -18.46 &  9.27 & 10.00 & -0.86 & -2.24 & -1.72 & 3322\nl
$20243-0226$ & -20.55 & 10.11 & 10.41 & -1.14 & -0.27 & -0.54 & 3332\nl
$20272-4738$ & -17.46 &  8.87 &  9.89 & -2.01 & -2.22 & -0.74 & 3332\nl
$20338+5958$ & -19.59 &  9.72 &  9.11 & -1.33 & -2.37 & -2.13 & 3322\nl
$20550+1655$ &\nodata &\nodata& 11.60 & -3.05 & -1.96 &  0.44 & 3322\nl
$20551-4250$ & -21.49 & 10.48 & 11.73 & -2.60 & -2.17 &  0.49 & 3332\nl
$21008-4347$ & -19.23 &  9.58 & 10.76 & -1.32 & -2.50 & -1.17 & 3332\nl
$21087+6557$ & -21.08 & 10.32 & 10.01 & -1.00 & -2.25 & -1.45 & 3322\nl
$21171-0859$ & -18.96 &  9.47 &  9.90 & -0.73 & -2.41 & -1.64 & 3332\nl
$21330-3846$ & -19.38 &  9.64 & 10.76 & -1.56 & -2.17 & -0.82 & 3332\nl
$21457-8145$ & -20.72 & 10.18 &  9.95 & -1.20 & -2.35 & -1.83 & 3332\nl
$23179+1657$ & -19.08 &  9.52 &  9.96 & -0.78 & -2.50 & -1.26 & 3332\nl
$23192-4245$ & -20.09 &  9.92 &  9.43 & -0.97 & -2.30 & -0.94 & 3332\nl
$23256+2315$ & -19.68 &  9.76 & 10.20 & -1.53 & -1.96 & -0.78 & 3332\nl
$23336+0152$ & -20.07 &  9.91 & 10.35 & -2.47 & -1.47 & -0.21 & 3232\nl
$23568+2028$ & -19.81 &  9.81 & 10.00 & -0.74 & -2.35 & -1.31 & 3332\nl
\enddata
\end{deluxetable}
\clearpage
\begin{deluxetable}{lccccccc}
\footnotesize
\tablecaption{Mean luminosities and ratios \label{tbl-3}}
\tablewidth{0pt}
\tablehead{
\colhead{Sample}&\colhead{N}& \colhead{$\log(\frac{L_B}{L_\odot})$}& \colhead{$\log(\frac{L_{IR}}{L_\odot})$} & \colhead{$\log(\frac{L_{IR}}{L_B})$} 
} 
\startdata
IRAS normal   &185&$9.9 \pm0.5$&$10.0 \pm0.5$&$0.12 \pm0.43$ \nl
PDS starbursts&182&$9.9 \pm0.4$&$10.3 \pm0.5$&$0.44 \pm0.48$ \nl
PDS Sy1       &31 &$10.3\pm0.4$&$10.4 \pm0.7$&$0.13 \pm0.49$ \nl
PDS Sy2       &62 &$10.1\pm0.5$&$10.4 \pm0.7$&$0.32 \pm0.49$ \nl
\enddata 
\end{deluxetable}

\clearpage
\begin{deluxetable}{llccccl}
\footnotesize
\tablecaption{Spectral characteristics of the galaxies\label{tbl-4}}
\tablewidth{0pt}
\tablehead{
\colhead{IRAS}&\colhead{Other name}& 
\colhead{$ \frac{ {\rm [OIII]}\lambda5007}{{\rm H}\beta} $} &
\colhead{$ \frac{ {\rm [NII]}\lambda6584}{ {\rm H} \alpha} $}  &
\colhead{$ \frac{ {\rm [SII]}\lambda(6717+6731)}{{\rm H}\alpha} $} & 
\colhead{$ \frac{ {\rm [OI]}\lambda6300}{{\rm H}\alpha} $}  &\colhead{Activity} \\
\colhead{}    &\colhead{}          &\colhead{}                                                 &
\colhead{}                                               &
\colhead{}                                               &
\colhead{}                                               &\colhead{Type}     \\ 
\colhead{}    &\colhead{}          &\colhead{}                                                 &
\colhead{}                                               &
\colhead{}                                               &
\colhead{}                                               &\colhead{}         
}
\startdata
$02395+3433$ & NGC1050    & 0.40& 1.47& 0.54& 0.05  & LINER      \nl
$03419+6756$ & IC0342     & 0.08& 0.43& 0.24& 0.01  & SBNG       \nl
$04435+1822$ & \nodata    & 2.37& 1.63& 0.28&\nodata& Sy2/LINER  \nl
$07107+3521$ & \nodata    & 4.10& 0.60& 0.31& 0.04  & Sy2        \nl
$07278-6728$ & IC2202     &11.72& 0.60& 1.21&\nodata& Sy2        \nl
$07568+1531$ & \nodata    &18.26& 1.34& 0.48& 0.12  & Sy2        \nl
$08339+6517X$&\nodata     & 1.65& 0.27& 0.25& 0.02  & SBNG       \nl
$10356+5345X$& NGC3310    & 1.94& 0.52& 0.40& 0.05  & SBNG/LINER \nl
$11257+5850X-E$& NGC3690  & 1.00& 0.25& 0.31& 0.04  & SBNG       \nl
$11257+5850X-W$& NGC3690  & 1.36& 0.44& 0.33& 0.05  & SBNG/LINER \nl
$12116+5448$ & MRK0201    & 1.13& 0.47& 0.26& 0.03  & SBNG       \nl
$12398-0641$ & NGC4628    & 8.36& 0.70& 0.46& 0.06  & Sy2        \nl
$12542-0815$ & NGC4818    & 0.14& 0.65& 0.19& 0.01  & SBNG/LINER \nl
$14454-4343$ & \nodata    &14.88& 1.04& 0.34& 0.09  & Sy2        \nl
$19265-4338$ &\nodata     & 6.48& 1.02& 0.72& 0.20  & Sy2        \nl
$20243-0226$ & IIZw083    & 3.43& 0.54& 0.44& 0.14  & Sy2        \nl
\enddata 
\end{deluxetable}

\end{document}